\documentclass[twocolumn]{aa}  
\usepackage{txfonts}
\usepackage{hyperref}
\usepackage[version=3]{mhchem}

\begin{document}

\title{Efficiency of radial transport of ices in protoplanetary disks probed with infrared observations: the case of \ce{CO2}} 
\author{Arthur D. Bosman \inst{1} \and Alexander G. G. M. Tielens \inst{1} \and Ewine F. van Dishoeck \inst{1,2} }
\institute{Leiden Observatory, Leiden University, PO Box 9513, 2300 RA Leiden, The Netherlands\\ \email{bosman@strw.leidenuniv.nl}\and
Max-Planck-Insitut f\"{u}r Extraterrestrische Physik, Gie{\ss}enbachstrasse 1, 85748 Garching, Germany}
\date{\today}

\abstract
{
Radial transport of icy solid material from the cold outer disk to the warm inner disk is thought to be important for planet formation. However, the efficiency at which this happens is currently unconstrained. Efficient radial transport of icy dust grains could significantly alter the composition of the gas in the inner disk, enhancing the gas-phase abundances of the major ice constituents such as \ce{H2O} and \ce{CO2}. }
{
Our aim is to model the gaseous \ce{CO2} abundance in the inner disk and use this to probe the efficiency of icy dust transport in a viscous disk. From the model predictions, infrared \ce{CO2} spectra are simulated and features that could be tracers of icy \ce{CO2}, and thus dust, radial transport efficiency are investigated.
}
{
We have developed a 1D viscous disk model that includes gas accretion and gas diffusion as well as a description for grain growth and grain transport. Sublimation and freeze-out of \ce{CO2} and \ce{H2O} has been included as well as a parametrisation of the \ce{CO2} chemistry. The thermo-chemical code DALI was used to model the mid-infrared spectrum of \ce{CO2}, as can be observed with \it{JWST}-MIRI. }
{
\ce{CO2} ice sublimating at the iceline increases the gaseous \ce{CO2} abundance to levels equal to the \ce{CO2} ice abundance of $\sim 10^{-5}$, which is three orders of magnitude more than the gaseous \ce{CO2} abundances of $\sim 10^{-8}$ observed by \textit{Spitzer}. Grain growth and radial drift increase the rate at which \ce{CO2} is transported over the iceline and thus the gaseous \ce{CO2} abundance, further exacerbating the problem. In the case without radial drift, a \ce{CO2} destruction rate of at least $10^{-11}$ s$^{-1}$ or a destruction timescale of at most 1000 yr is needed to reconcile model prediction with observations. This rate is at least two orders of magnitude higher than the fastest destruction rate included in chemical databases. A range of potential physical mechanisms to explain the low observed \ce{CO2} abundances are discussed.
}
{
We conclude that transport processes in disks can have
profound effects on the abundances of species in the inner disk such
as \ce{CO2}. The discrepancy between our model and observations either suggests
frequent shocks in the inner 10 AU that destroy \ce{CO2}, or that the abundant midplane \ce{CO2}
is hidden from our view by an optically thick column of low abundance
\ce{CO2} due to strong UV and/or X-rays in the surface layers. Modelling
and observations of other molecules, such as \ce{CH4} or \ce{NH3}, can give 
further handles on the rate of mass transport.
}

\keywords{Protoplanetary disks -- astrochemistry -- accretion, accretion disks -- methods: numerical}
\titlerunning{Efficiency of radial transport of ices: the case of \ce{CO2}}
\maketitle 
\section{Introduction}

To date, a few thousand planetary systems have been found\footnote{\url{exoplanets.org} as of 28 Nov 2017}.  Most of them have system architectures that are very different from our own solar system \citep{Madhusudhan2014} and explaining the large variety of systems is a challenge for current planet formation theories \citep[see, e.g.][and references therein]{Morbidelli2016}. Thus, the birth environment of planets -- protoplanetary disks -- are an active area of study. A young stellar system inherits small dust grains from the interstellar medium. In regions with high densities and low turbulence, the grains start to coagulate. In the midplane of protoplanetary disks, where densities are higher than $10^8$ cm$^{-3}$, grain growth can really take off. Grain growth and the interactions of these grown particles with the gaseous disk are of special interest to planet formation \citep[see, e.g.][]{Weidenschilling1977,Lambrechts2012,Testi2014}. The growth of dust grains to comets and planets is far from straightforward, however.

Pebbles, that is, particles that are large enough to slightly decouple
from the gas, have been invoked to assist the formation of planets in
different ways \citep{Johansen2017}. They are not
supported by the pressure gradient from the gas, but they are subject
to gas drag. As a result pebbles drift on a time-scale that is an
order of magnitude faster than the gas depletion time-scale. This flow
of pebbles, if intercepted or stopped, can help with planet
formation. The accretion of pebbles onto forming giant planetary cores
should help these cores grow beyond their classical isolation mass
\citep{Ormel2010,Lambrechts2012} while the interactions of gas and
pebbles near the inner edge of the disk can help with the formation of
ultra compact planetary systems as found by the {\it Kepler} satellite
\citep{Tan2016,Ormel2017}. Efficient creation and redistribution of
pebbles would lead to quick depletion of the solid content of disks
increasing their gas-to-dust ratios by one to two orders of
magnitude in 1 Myr
\citep[e.g.][]{Ciesla2006,Brauer2008,Birnstiel2010}.

Observations of disks do not show evidence of strong dust
depletion. The three disks that have far-infrared measurements of the HD molecule
to constrain the gas content show gas-to-dust ratios smaller than 200
\citep{Bergin2013,McClure2016,Trapman2017}, including the 10 Myr old
disk TW Hya \citep{Debes2013}. Gas-to-dust ratios found from gas mass
estimates using \ce{CO} line fluxes are inconsistent with strong
dust depletion as well \citep[e.g.][]{Ansdell2016,
  Miotello2017}. \cite{Manara2016b} also argue from the observed
relation between accretion rates and dust masses, that the gas-to-dust
ratio in the 2--3 Myr old Lupus region should be close to 100.

It is thus of paramount importance that a way is found to quantify the
inwards mass flux of solid material due to radial drift from
observations. This is especially important for the inner ($< 10$
AU) regions of protoplanetary disks. Here we propose to look for a
signature of efficient radial drift in protoplanetary disks through
molecules that are a major component of icy planetesimals.

Radial drift is expected to transport large amounts of ice over the
iceline, depositing a certain species in the gas phase just inside and
ice just outside the iceline
\citep{Stevenson1988,Piso2015,Oberg2016}. This has been modelled in
detail for the water iceline by \cite{Ciesla2006} and
\cite{Schoonenberg2017}, for the \ce{CO} iceline by
\cite{Stammler2017} and in general for the \ce{H2O}, \ce{CO} and
\ce{CO2} icelines by \cite{Booth2017}. In all cases the gaseous
abundance of a molecule in the ice is enhanced inside of the iceline
as long as there is an influx of drifting particles. The absolute
value and width of the enhancement depend on the mass influx of ice
and the strength of viscous mixing. Such an enhancement may be seen
directly with observations of molecular lines. Out of the three most
abundant ice species (\ce{CO}, \ce{H2O} and \ce{CO2}), \ce{CO2} is the
most promising candidate for a study of this nature. Both \ce{CO} and
\ce{H2O} are expected to be highly abundant in the inner disk based on
chemical models \citep[see,
e.g.][]{Aikawa1999,Markwick2002,Walsh2015,Eistrup2016}. As such any
effect of radial transport of icy material will be difficult to
observe. \ce{CO2} is expected to be abundant in outer disk ices
\citep[with abundances around $10^{-5}$ if inherited from the cloud or
produced in situ in the
disk,][]{Pontoppidan2008,Boogert2015,LeRoy2015,Drozdovskaya2016}, but
it is far less abundant in the gas in the inner regions of the disk
\citep[with abundances around
$10^{-8}$,][]{Pontoppidan2010,Bosman2017}. This gives a leverage of
three orders of magnitude to see effects from drifting icy pebbles.

Models by \cite{Cyr1998,AliDib2014}, for example, predict depletion of
volatiles in the inner disk. In these models, all volatiles are locked
up outside of the iceline in stationary solids. In \cite{AliDib2014} this
effect is strengthened by the assumption that the gas and the small dust in the disk midplane are moving radially outwards such as proposed by \cite{Takeuchi&Lin2002}. Together this leads to very low inner disk \ce{H2O}
abundances in their models. \ce{CO2} is not included in these models,
but the process for \ce{H2O} would also work for \ce{CO2}, but
slightly slower, as the \ce{CO2} iceline is slightly further
out. However, these models do not include the diffusion of small
grains through the disk, which could resupply the inner disk with
volatiles at a higher rate than that caused by the radial drift of large
grains.
 
\cite{Bosman2017} showed that an enhancement of \ce{CO2} near
the iceline would be observable in the \ce{^{13}CO2} mid infrared
spectrum, if that material were mixed up to the upper layer. Here we
continue on this line of research by constraining the maximal mass
transport rate across the iceline and by investigating the shape and
amplitude of a possible \ce{CO2} abundance enhancement near the
iceline due to this mass transport. 

To constrain the mass transport we have build a model along the same
lines as \cite{Ciesla2006} and \cite{Booth2017} except that we do not
include planet formation processes. As in \cite{Booth2017} we use the
dust evolution prescription from \cite{Birnstiel2012}. The focus is on
the chemistry within the \ce{CO2} iceline to make predictions on the
\ce{CO2} content of the inner disk. In contrast with
\cite{Schoonenberg2017} a global disk model is used to maintain
consistency between the transported mass and observed outer disk
masses. Chemical studies of the gas in the inner disk have been
presented by, for example \cite{Agundez2008}, \cite{Eistrup2016},
\cite{Walsh2015} but transport processes are not included in these
studies. \cite{Cridland2017} use an evolving disk, including grain
growth and transport, coupled with a full chemical model in their
planet formation models. However, they do not include transport of ice
and gas species due to the various physical evolution processes.

\S 2 presents the details of our physical model, whereas \S 3
discusses various midplane chemical processes involving CO$_2$ and our
method for simulating infrared spectra. \S 4 presents the model
results for a range of model assumptions and parameters. One common
outcome of all of these models is that the CO$_2$ abundance in the
inner disk is very high, orders of magnitude more than observed,
making it difficult to quantify mass transport. \S 5 discusses
possible ways to mitigate this discrepancy and the implications for
the physical and chemical structure of disks, suggesting JWST-MIRI
observations of $^{13}$CO$_2$ that can test them.

\section{Physical model}
\subsection{Gas dynamics}
To investigate the effect of drifting pebbles on inner disk gas-phase abundances we build a 1-D dynamic model. This model starts with an $\alpha$-disk model \citep{Shakura1973}. The evolution of the surface density of gas $\Sigma_\mathrm{gas}(t,r)$ can be described as:
\begin{equation}
\label{eq:viscous}
\frac{\partial\Sigma_{\mathrm{gas}}}{\partial t} = \frac{3}{r}\frac{\partial}{\partial r} \left[r^{1/2}\frac{\partial}{\partial r} \left(\frac{\alpha c_s^2}{\Omega} \Sigma_{\mathrm{gas}} r^{1/2} \right) \right],
\end{equation} 

where $r$ is the distance to the star, $t$ is time, $\alpha$ is the dimensionless Sakura-Sunyaev parameter, $\Omega$ is the local Keplerian frequency and $c_s = \sqrt{k_b T/\left(\mu \right)}$ is the local sound speed, with $k_b$ the Boltzmann constant, T the local temperature and $\mu$ the mean molecular mass which is taken to be 2.6 amu. $\alpha c_s^2/\Omega$ is also equal to $\nu_{\mathrm{turb}}$, the (turbulent) gas viscosity. The viscosity, or the resistance of the gas to shear, is responsible for the exchange of angular momentum. The origin of the viscosity is generally assumed to be turbulence stirred up by the Magneto-Rotational Instability (MRI) \citep{Turner2014} although Eq.~(\ref{eq:viscous}) is in principle agnostic to the origin of the viscosity as long as the correct $\alpha$ value is included. For the rest of the paper, we assume that the viscosity originates from turbulence and that the process responsible for the viscous evolution is also the dominant process in mixing constituents of the disk radially. 

The evolution of the surface density of a trace quantity has two main components. First, all gaseous constituents are moving with the flow of the gas. This advection is governed by:

\begin{equation}
\label{eq:advect}
\frac{\partial \Sigma_{x, \mathrm{gas}}}{\partial t} = \frac{1}{r}\frac{\partial}{\partial r} \left( F_{\mathrm{gas}} r  \frac{\Sigma_{x, \mathrm{gas}}}{\Sigma_{\mathrm{gas}}} \right), 
\end{equation}
where $F_{\mathrm{gas}}$ is the total radial flux, which is related to the radial velocity of the gas due to viscous accretion, $u_\mathrm{gas}$, and is given by: 
\begin{equation}
\label{eq:flux}
F_{\mathrm{gas}}= \Sigma_{\mathrm{gas}} u_\mathrm{gas} = \frac{3}{\sqrt{r}} \frac{\partial}{\partial r} \nu \Sigma_\mathrm{gas}\sqrt{r} .
\end{equation}

Second, the gas is also mixed by the turbulence, smoothing out variations in abundance. This diffusion can be written as \citep{ClarkePringle1988,Desch2017}
\begin{equation}
\label{eq:diff}
\frac{\partial \Sigma_{x, \mathrm{gas}}}{\partial t} = \frac{1}{r} \frac{\partial}{\partial r} \left( r D_{x} \Sigma_{\mathrm{gas}} \frac{\partial}{\partial r}\left(\frac{\Sigma_{x, \mathrm{gas}}}{\Sigma_{\mathrm{gas}} }\right) \right),
\end{equation}
where $D_x$ is the gas mass diffusion coefficient. The diffusivity is related to the viscosity by: 
\begin{equation}
\mathrm{Sc} = \frac{\nu_\mathrm{gas}}{D_x},
\end{equation}
with $\mathrm{Sc}$ the Schmidt number. For turbulent diffusion in a viscous disk it is expected to be of order unity. As such, $\mathrm{Sc}=1$ is assumed for all gaseous components. 

Advection and diffusion are both effective in changing the abundance of a trace species if an abundance gradient exists. Diffusion is most effective in the presence of strong abundance gradients and strong changes in the abundance gradient, such as near the iceline. At the icelines the diffusion of a trace species will generally dominate over the advection due to gas flow.

\subsection{Dust growth and dynamics}

The dynamics of dust grains are strongly dependent on the grain size. A grain with a large surface area relative to its mass is well coupled to the gas and will not act significantly different from a molecule in the gas. Solid bodies with a very small surface area relative to its mass, for example, planetesimals, will not be influenced by the gas pressure or turbulence, their motions are then completely governed by gravitational interactions. Dust particles with sizes between these extremes will be influenced by the presence of the gas in various ways. To quantify these regimes it is useful to consider a quantity known as the Stokes number: The Stokes number is, assuming Epstein drag and spherical particles in a vertically hydrostatic disk, given by \citep{Weidenschilling1977,Birnstiel2010}: 
\begin{equation}
\mathrm{St} = \frac{a_\mathrm{grain}\rho_{s}}{\Sigma_{\mathrm{gas}}}\frac{\pi}{2}.
\end{equation} 
Particles with a very small Stokes number ($\ll1$) are very well coupled to the gas and the gas pressure gradients and particles with very large Stokes number ($\gg 1$) are decoupled from the gas.

The coupling between gas and dust determines both the diffusivity of the dust, that is, how well the dust mixed due to turbulent motions of the disk as well as advection of the dust due to bulk flows of the gas. \cite{Youdin2007} derived that the diffusivity $D_\mathrm{dust}$ of a particle with a certain Stokes number can be related to the gas diffusivity by:
\begin{equation}
\label{eq:dust_diff}
D_\mathrm{dust} = \frac{D_\mathrm{gas}}{1+ \mathrm{St}^2}.
\end{equation}
Similarly the advection speed of dust due to gas advection can be given by:
\begin{equation}
u_\mathrm{dust,adv} = \frac{u_\mathrm{gas} }{1+ \mathrm{St}^2}.
\end{equation}

When particles are not completely coupled to the gas they are also no longer completely supported by the gas pressure gradient. The gas pressure gradient, from high temperature, high density material close to the star, to low temperature, low density material far away from the star provides an outwards force, such that the velocity with which the gas has a stable orbit around the star is lower than the Keplerian velocity. Particles that start to decouple from the gas thus also need to speed up relative to the gas to stay in a circular orbit. This induces a velocity difference between the gas and the dust particles. The velocity difference between the gas velocity and a Keplerian orbital velocity is given by:
\begin{equation}
\Delta u = \Omega r - \sqrt{\Omega^2r^2 - \frac{r}{\rho_\mathrm{gas}}\frac{\partial P}{\partial r}},
\end{equation}
with $P$ the pressure of the gas. We note that in the case of a positive pressure gradient, the gas will be moving faster than the Keplerian velocity

As a result of this velocity difference the particles are subjected to a drag force, which, in the case of a negative pressure gradient, removes angular momentum from the particles. 
The loss of angular momentum results in an inwards spiral of the dust particles. This process is known as radial drift. The maximal drift velocity can be given as \citep{Weidenschilling1977}:
\begin{equation}
\label{eq:driftvel}
u_\mathrm{drift} = \frac{2 \Delta u}{\mathrm{St} + \mathrm{St}^{-1}}.
\end{equation}
The drift velocity will thus be maximal for particles with a Stokes number of unity. Drift velocities of $\sim 1\%$ of the orbital velocity are typical for particles with a Stokes number of unity. 

The dynamics of dust are thus intrinsically linked to the size, or rather size distribution of the dust particles. The dust size distribution is set by the competition of coagulation and fragmentation\footnote{Cratering and bouncing are neglected for simplicity}. When two particles collide in the gas they can either coagulate, that is, the particles stick together and continue on as a single larger particle, or they can fragment, the particles break apart into many smaller particles. The outcome of the collision depends on the relative velocity of the particles and their composition. At low velocities particles are expected to stick, while at high velocities collisions lead to fragmentation. The velocity that sets the boundary between the two outcomes is called the fragmentation velocity. For pure silicate aggregates the fragmentation velocities are low (1 m s$^{-1}$) while aggregates with a coating of water ice can remain intact in collisions with relative velocities up to a order of magnitude higher \citep{Blum2008,Gundlach2015}. 

Dust size distributions resulting from the coagulation and fragmentation processes can generally not be computed analytically. Calculations have been done for both static and dynamic disks \citep{Brauer2008,Birnstiel2010,Krijt2016}. These models are very numerically intensive and do not lend themselves to the inclusion of additional physics and chemistry nor to large parameter studies. As such we will use the simplified dust evolution prescription from \cite{Birnstiel2012}. This prescription has been benchmarked against models with a more complete grain growth and dust dynamics prescription from \cite{Brauer2008}. The prescription only tracks the ends of the dust size distribution, the smallest grains of set size and the largest grains at a location in the disk of a variable size. As this prescription is a key part of the model we will reiterate some of the key arguments, equations and assumption of this prescription here, for a more complete explanation, see \cite{Birnstiel2012}.

The prescription assumes that the dust size distribution can be in one of three stages at any point in the disk. Either the largest particles are still growing, the largest particle size is limited by radial drift, or the largest particle size is limited by fragmentation. In the first and second case, the size distribution is strongly biased towards larger sizes. In the final case the size distribution is a bit flatter \citep[see,][]{Brauer2008}. The size distribution in all cases is parametrised by a minimal grain size, the monomer size, a maximal grain size, which depends on the local conditions, and the fraction of mass in the large grains. 

From physical considerations one can write a maximal expected size of the particles due to different processes. Growth by coagulation is limited by the amount of collisions and thus by the amount of time that has passed. The largest grain expected in a size distribution at a given time is given by:
\begin{equation}
a_\mathrm{grow}(t) = a_\mathrm{mono}\exp\left[{\frac{t-t_0}{\tau_\mathrm{grow}}}\right],
\end{equation}
where $a_\mathrm{mono}$ is the monomer size, $t_0$ is the starting time, $t$ is the current time and $\tau_\mathrm{grow}$ is the local growth time scale, given by:
\begin{equation}
\tau_\mathrm{grow} = \frac{\Sigma_\mathrm{gas}}{\Sigma_\mathrm{dust}\Omega}.
\end{equation}
Radial drift moves the particles inwards: the larger the particles, the faster the drift. There is thus a size at which particles are removed faster due to radial drift than they are replenished by grain growth, limiting the maximal size of particles. This maximal size is given by:
\begin{equation}
a_\mathrm{drift} = f_d \frac{8 \Sigma_{\mathrm{dust}}}{\rho_s}\frac{\pi r^2 \Omega^2}{c_s^2}\gamma^{-1},
\end{equation}
with $f_d$ a numerical factor, $\rho_s$ is the density of the grains and $\gamma$ is the absolute power law slope of the gas pressure:
\begin{equation}
\gamma = \left|\frac{\mathrm{d}\ln P_\mathrm{gas}}{\mathrm{d} \ln r}\right|.
\end{equation}
As mentioned before, particles that collide with high velocities are expected to fragment. For this model we consider two sources of relative velocities. One source of fragmentation is differential velocities due to turbulence \citep[see,][for a more complete explanation]{Ormel2007}. This limiting size is given by: 
\begin{equation}
a_\mathrm{frag} = f_f \frac{2}{3\pi}\frac{\Sigma_\mathrm{gas}}{\rho_s \alpha}\frac{u_f^2}{c_s^2},
\end{equation}
where $f_f$ is a numerical fine tuning parameter and $u_f$ is the fragmentation velocity. The other source of fragmentation considered is different velocities due to different radial drift speeds. The limiting size for this process is given by:
\begin{equation}
a_\mathrm{df} = \frac{u_f r \Omega}{ c_s^2(1-N)}\frac{4\Sigma_\mathrm{gas}}{\rho_s} \gamma^{-1},
\end{equation}
$N$ is the average Stokes number fraction between two colliding grains, here we use $N = 0.5$.

The size distribution at a location in the disk spans from the monomer size ($a_\mathrm{mono}$) to the smallest of the four limiting sizes above. In the model the grains size distribution is split into `small' and `large' grains. The mass fraction of the large grains depends on the process limiting the size: if the grain size is limited by fragmentation ($\min(a_\mathrm{frag}, a_\mathrm{df}) < a_\mathrm{drift}$), the fraction of mass in large grains ($f_{m,\mathrm{\,frag}}$) is assumed to be 75\%. If the grain size is limited by radial drift ($a_\mathrm{drift} < \min(a_\mathrm{frag}, a_\mathrm{df}$)), the fraction of mass in large grains ($f_{m,\mathrm{\,drift}}$) is assumed to be 97\%. These fractions were determined by \cite{Birnstiel2012} by matching the simplified model to more complete grain-growth models.

Using these considerations the advection-diffusion equation for the dynamics of the dust can be rewritten:

\begin{equation}
\begin{split}
\frac{\partial \Sigma_{\mathrm{dust}}}{\partial t} =  \frac{1}{r}\frac{\partial}{\partial r} \left[\vphantom{\frac12}\left( u_{\mathrm{dust,\,s\,}} r  (1-f_m)\Sigma_{\mathrm{dust}} + u_{\mathrm{dust,\,l\,}} r  f_m\Sigma_{\mathrm{dust}}\right) \right.\\ 
+r \Sigma_{\mathrm{gas}} \left( D_\mathrm{dust,\,s\,} \frac{\partial}{\partial r}\left(\frac{(1-f_m)\Sigma_{\mathrm{dust}}}{\Sigma_{\mathrm{gas}} }\right) \right.\\
+ \left.\left. D_{\mathrm{dust,\,l\,}} \frac{\partial}{\partial r}\left(\frac{f_m\Sigma_{\mathrm{dust}}}{\Sigma_{\mathrm{gas}} }\right) \right) \right],
\end{split}
\end{equation}
where $D_\mathrm{dust,\,x\,} = D_\mathrm{gas}/{(1+\mathrm{St}_\mathrm{x}^2)}$,  $u_{\mathrm{dust,\,x\,}} = {u_\mathrm{gas}}/{(1 + \mathrm{St}_\mathrm{x}^2)}  + u_\mathrm{drift,\,x}$ where $u_\mathrm{drift}$ is the velocity due to radial drift (Eq.~\ref{eq:driftvel}). Here it is assumed that the large grains never get a Stokes number larger than unity, which holds for the models presented here.

The final part of the dynamics concerns the ice on the dust. The ice moves with the dust, so both the large scale movements as well as the mixing diffusion of dust must be taken into account. For the model we assume that the ice is distributed over the grains according to the mass of the grains. This means that if large grains have a mass fraction $f_m$, then the large grains will have the same mass fraction of $f_m$ of the available ice. This is a reasonable assumption, as long as the grain size distribution is in fragmentation equilibrium and the largest grains are transported on a timescale longer than the local growth timescale. The algorithm used here forces the latter condition to be true everywhere in the disk, while the former condition is generally true for the \ce{CO2} and \ce{H2O} icelines \citep{Brauer2008}, but not for the \ce{CO} iceline \citep{Stammler2017}.
 With this assumption the advection-diffusion equation for the ice can be written as:

\begin{equation}
\begin{split}
\frac{\partial \Sigma_{\mathrm{ice}, x}}{\partial t} = 
\frac{1}{r} \frac{\partial}{\partial r} r&\left(F_{\mathrm{dust}} \frac{\Sigma_{\mathrm{ice}, x}}{\Sigma_\mathrm{dust}}  \right.\\
 &\left.+ \Sigma_{\mathrm{dust}} \left( D_\mathrm{dust,\,s\,}(1-f_m) +D_\mathrm{dust,\,l\,}f_m  \right) \frac{\partial}{\partial r}\frac{\Sigma_{\mathrm{ice,x}}}{\Sigma_{\mathrm{dust}} }\right).
\end{split}
\end{equation}

Here $F_{\mathrm{dust}}$ is the radial flux of dust, this is given by:
\begin{equation}
\begin{split}
F_{\mathrm{dust}} &=  \Sigma_{\mathrm{dust}}\left(u_{\mathrm{dust,\,s\,}}  (1-f_m) + u_{\mathrm{dust,\,l\,}} f_m\right) \\ 
&+ \Sigma_{\mathrm{dust}} \left( D_\mathrm{dust,\,s\,} \frac{\partial}{\partial r}\left(\frac{(1-f_m)\Sigma_{\mathrm{dust}}}{\Sigma_{\mathrm{gas}} }\right) + D_{\mathrm{dust,\,l\,}} \frac{\partial}{\partial r}\left(\frac{f_m\Sigma_{\mathrm{dust}}}{\Sigma_{\mathrm{gas}} }\right) \right).
\end{split}
\end{equation}

\subsection{Model parameters}

For our study, we pick a disk structure that should be representative of a young disk around a one solar mass star. The initial surface density structure is given by:
\begin{equation}
\Sigma_{\mathrm{gas}}(r) = \Sigma_{1 \mathrm{AU}} \left(\frac{r}{1 \mathrm{AU}}\right)^{-p}
 \exp\left[\left(-\frac{r}{r_\mathrm{taper}}\right)^{2-p}\right]
\end{equation}
where $p$ controls the steepness of the surface density profile, $r_\mathrm{taper}$ the extent of the disk and $\Sigma_{1 \mathrm{AU}}$ sets the normalisation of the surface density profile. For our models we use $p = 1$ and $r_\mathrm{taper} = 40$ AU. The temperature profile is given by a simple power law: 
\begin{equation}
T(r) = T_{1 \mathrm{AU}} \left( \frac{r}{1 \mathrm{AU}}\right)^{-q}.
\end{equation}
The disk is assumed to be viscous with a constant $\alpha$, as such there is a constraint on the power law slopes $p$ and $q$, if we want the gas surface density to be a self similar solution to Eq.~\ref{eq:viscous}, it is required to have $p + q = 3/2$ \citep{Hartmann1998}. 

To calculate the volume densities, a vertical Gaussian distribution of gas with a scale height $H_p(r) = h_p r$, with $h_p = 0.05$, is used.

The disk is populated with particles of $0.1 \mu$m at the start of the model, this is also the size of the small dust in the model. The gas-to-dust ratio is taken to be 100 over the entire disk. The density of the grains is assumed to be 2.5 g cm$^{-3}$ and grains are assumed to be solid spheres. 

The molecules are initially distributed through simple step functions. These step functions have a characteristic radius `iceline' which differentiates between the inner disk and the outer disk. Within this radius the gas phase abundance of the molecule is initialised, outside of this radius the solid phase abundance is initialised. Water is distributed equally through the disk with an abundance of $1.2 \times 10^{-4}$, whereas \ce{CO2} is initially depleted in the inner disk with an abundance of $10^{-8}$ \citep{Pontoppidan2011,Bosman2017} and has a high abundance in the outer disk of $4 \times 10^{-5}$. This ice abundance of \ce{CO2} is motivated by the ISM ice abundance \citep{Pontoppidan2008, Boogert2015}, cometary abundances \citep{LeRoy2015} and chemical models of disk formation \citep{Drozdovskaya2016}.

A summary of the initial conditions, fixed and variable parameters is given in Table~\ref{tab:initialconditions}.

\begin{table}
\centering
\caption{\label{tab:initialconditions} Initial conditions, fixed parameters and variables}
\small
\begin{tabular}{lll}
\hline
\hline
Description & symbol & value \\ 
\hline
\multicolumn{3}{c}{Initial conditions and fixed parameters} \\
\hline
\multicolumn{3}{c}{Disk physical structure} \\
\hline
Central stellar mass & $M_\star$ & 1 M$_\odot$\\
Surface density at 1 AU & $\Sigma_{1 \mathrm{AU}}$ & 15000 kg m$^{-2}$ \\
Temperature at 1 AU & $ T_{1 \mathrm{AU}}$ & 300 K \\
Exponential taper radius & $r_\mathrm{taper}$ & 40 AU \\
$\Sigma$ density power law slope & $p$ & 1 \\
$T$ power law slope  & $q$ & 0.5\\
Total initial disk mass & $M_{\mathrm{gas},0}$ & 0.02 M$_\odot$\\
Disk FWHM angle & $h_\mathrm{FWHM}$ & 0.05 rad\\
\hline
\multicolumn{3}{c}{Dust properties} \\
\hline
Gas-to-dust ratio & g/d & 100 \\
Monomer size & $a_\mathrm{small}$ & 0.1 $\mu$m\\
Grain density & $\rho_s$ & 2.5 g cm$^{-3}$\\
Drift lim. large grain frac. & $f_{m,\mathrm{\,drift}}$ & 0.97\\
Frag. lim. large grain frac. & $f_{m,\mathrm{\,frag}}$ & 0.75\\
Num. factor fragmentation size & $f_f$ & 0.37 \\
Num. factor drift size & $f_d$ & 0.55\\ 
\hline
\multicolumn{3}{c}{Chemical parameters} \\
\hline
Sticking fraction & $f_s$ & 1.0 \\
Cosmic ray ionisation rate & $\zeta_{\ce{H2}}$ & $10^{-17}$ s$^{-1}$\\
Number of active ice layers & $N_\mathrm{act}$ & 2 \\ 
Step radius water & $r_\mathrm{step, \ce{H2O}}$ & 3 AU \\
Inner disk \ce{H2O} abundance & $x_\mathrm{in, \ce{H2O}}$ & $1.2\times 10^{-4}$ \\
Outer disk \ce{H2O} abundance & $x_\mathrm{out, \ce{H2O}}$ & $1.2\times 10^{-4}$ \\
\ce{H2O} binding energy   & $E_\mathrm{bind, \ce{H2O}} $ & 5600 K \\
\ce{H2O} desorption prefactor & $p_{\ce{H2O}}$& $10^{30}$ cm$^{-2}$ s$^{-1}$\\ 
\ce{H2O} CR destruction efficiency & $k_{\ce{H2O}}/\zeta_{\ce{H2}}$ & 1800\\
Step radius \ce{CO2} & $r_\mathrm{step, \ce{CO2}}$ & 10 AU \\
Inner disk \ce{CO2} abundance & $x_\mathrm{in, \ce{CO2}}$ & $1\times 10^{-8}$ \\
Outer disk \ce{CO2} abundance & $x_\mathrm{out, \ce{CO2}}$ & $4\times 10^{-5}$ \\
\ce{CO2} binding energy   & $E_\mathrm{bind, \ce{CO2}} $ & 2900 K \\
\ce{CO2} desorption prefactor & $p_{\ce{CO2}} $& $9.3 \times 10^{26}$\,cm$^{-2}$\,s$^{-1}$\\ 

\hline
\multicolumn{3}{c}{Variables} \\
\hline
Viscosity parameter & $\alpha$ & \\
Fragmentation velocity & $u_\mathrm{f}$ & \\
\ce{CO2} destruction rate & $R_{\mathrm{destr, }\ce{CO2}}$& \\
\hline
\end{tabular}
\end{table}

\subsection{Boundary conditions}

Due to finite computational capabilities, the calculation domain of the model needs to be limited. The assumptions made for the edges can have significant influences on the model results. For the inner edge, it is assumed that gas leaves the disk with an accretion rate as assumed from a self similar solution ($p + q = 3/2$) according to the viscosity and surface density at the inner edge. The accretion rate is given by: 
\begin{equation}
\dot{M} = 3\pi\Sigma_\mathrm{gas}\nu_\mathrm{gas}.
\end{equation}
All gas constituents advect with the gas over the inner boundary or, in the case of large grains, drift over the boundary. Diffusion over the inner boundary is not possible.
For the outer boundary, the assumption is made that the surface density of gas outside the domain is zero. Again viscous evolution or an advection process can remove grains and other gas constituents from the computational domain.

\section{\label{sec:chemproc}Chemical processes}

\subsection{Freeze-out and sublimation}
Freeze-out and sublimation determine the fraction of a molecule that is in the gas phase and the fraction that is locked up on the grains. Freeze-out of a molecule, or a molecule's accretion onto a grain is given by the collision rate of a molecule with the grain surface times the sticking fraction, $f_s$, which is assumed to be unity:
\begin{equation}
R_\mathrm{acc,x} = f_s \sigma_{\mathrm{dust}} n_\mathrm{grain} v_\mathrm{therm,x}, 
\end{equation}
where $\sigma_{\mathrm{dust}}$ is the average dust surface area, $n_\mathrm{grain}$ is the number density of grains and the thermal velocity $v_\mathrm{therm,x} = \sqrt{8kT / \pi m_x}$, with $k$ the Boltzmann constant, $T$ the temperature and $m_x$ the mass of the molecule. Molecules that are frozen-out on the grain can sublimate or desorb. For a grain covered with many monolayers of ice, the rate per unit volume for this process is given by:
\begin{equation}
f_\mathrm{des,x} = R_\mathrm{des,x} n_\mathrm{x,ice} = p_x \sigma_{\mathrm{dust}} n_\mathrm{grain} N_\mathrm{act} \exp\left[-\frac{E_\mathrm{bind}}{kT}\right], 
\end{equation}
where $p_x$ is the zeroth-order `prefactor' encoding the frequency of desorption attempts per unit surface area, $\sigma_{\mathrm{dust}}$ is the surface area per grain, $n_\mathrm{grain}$ is the number density of grains. $N_{\mathrm{act}}$ is the number of active layers, that is, the number of layers that can participate in the sublimation, $N_\mathrm{act} = 2$ is used, $T$ is the dust temperature, which we take equal to the gas temperature. For mixed ices this rate can be modified by a covering fraction $\chi_x = n_{\mathrm{ice,\,}x}/ \sum_xn_{\mathrm{ice,\,}x} $, however, this is neglected here. We note that $f_\mathrm{des,x}$ in its current form is independent of the amount of molecules frozen out on the dust grains. Using these rates we get the following differential equation:
\begin{equation}
\frac{\partial n_\mathrm{x,gas}}{\partial t} = -R_\mathrm{acc,x} n_\mathrm{x,gas} + f_\mathrm{des,x},
\end{equation}
which has an analytical solution:
\begin{equation}
\begin{split}
n_\mathrm{x,gas}(t) &= \min\left[\vphantom{\frac12}n_\mathrm{x,tot},\right.\\
& \left. \left(n_\mathrm{x,gas}(t_0) - \frac{f_\mathrm{des,x}}{R_\mathrm{acc,x}}\right)\exp\left(-R_\mathrm{acc,x}(t-t_0)\right) + \frac{f_\mathrm{des,x}}{R_\mathrm{acc,x}}\right].
\end{split}
\end{equation}
where $n_\mathrm{x,tot}$ is the total number density of a molecule (gas and ice). The number density of ice is given by: 
\begin{equation}
n_\mathrm{x,ice}(t) = n_\mathrm{x,tot} - n_\mathrm{x,gas}(t).
\end{equation}

The ice line temperature, defined as the temperature for which $n_\mathrm{x,gas}=n_\mathrm{x,dust}$, that is, when freeze-out and sublimation balance, depends on the total number density of the molecule considered. At lower total molecule number densities, the iceline will be at lower temperatures. 

For \ce{CO2} we use a binding energy of 2900 K, representative for \ce{CO2} mixed with water \citep{Sandford1990, Collings2004}. More recent measurements have suggested that the binding energy is lower, around 2300 K \citep{Noble2012}. Using the lower binding energy moves the \ce{CO2} iceline further out from 90 to 80 K or from 6 to 8 AU in our standard model. The prefactor, $p_{x,\,\ce{CO2}}$, of $9.3 \times 10^{26}$\,cm$^{-2}$\,s$^{-1}$ from \citep{Noble2012} is used. The change in iceline location has a minimal effect on the evolution of the abundance profiles. Tthe mixing time-scale becomes longer at larger radii which would increase mixing times. For \ce{H2O} a binding energy of 5600 K and prefactor, $p_{x,\,\ce{H2O}}$, of $10^{30}$\,cm$^{-2}$\,s$^{-1}$ \citep{Fraser2001}.

\subsection{Midplane formation and destruction processes}
\label{ssc:FormDestr}
Radial drift and radial diffusion and advection will quickly move part of the outer disk \ce{CO2} ice reservoir into the inner disk, enhancing the inner disk abundances. To get a good measure of the amount of \ce{CO2} in the inner disk it is necessary to also take into account the processes that form and destroy \ce{CO2} in gas and ice. The density is highest near the mid-plane, as such this is where formation and destruction process are expected to be most relevant for the bulk of the \ce{CO2}. However, in the less dense upper layers, there are UV-photons that can dissociate and ionise molecules, possibly influencing the overall abundance of \ce{CO2}. 

\subsubsection{Gas-phase formation of \ce{CO2}}
The formation of \ce{CO2} in the inner disk mainly goes through the warm gas-phase route. Here \ce{CO2} forms through the reaction \ce{CO + OH -> CO2 + H}. The reaction has a slight activation barrier of 176 K \citep{Smith2004}. The parent molecule \ce{CO} is very stable and is expected to be present at high abundances in the inner disk ($ 10^{-4}$) \citep{Walsh2015}. The \ce{OH} radical is expected to be less abundant, and it is the fate of this radical that determines the total production rate of \ce{CO2}. \ce{OH} is formed either directly from \ce{H2O} photodissociation \citep{Heays2017}, or by hydrogenation of atomic oxygen, \ce{O + H2 -> OH +H}, in a reaction that has an activation barrier of 3150 K \citep{Baulch1992}. The atomic oxygen itself also has to be liberated from, in this case, either \ce{CO}, \ce{CO2} or \ce{H2O} by X-rays or UV-photons. The production rate of \ce{CO2} is thus severely limited if there is no strong radiation field present to release oxygen from one of the major carriers. 

\begin{figure}
\centering
\includegraphics[width=\hsize]{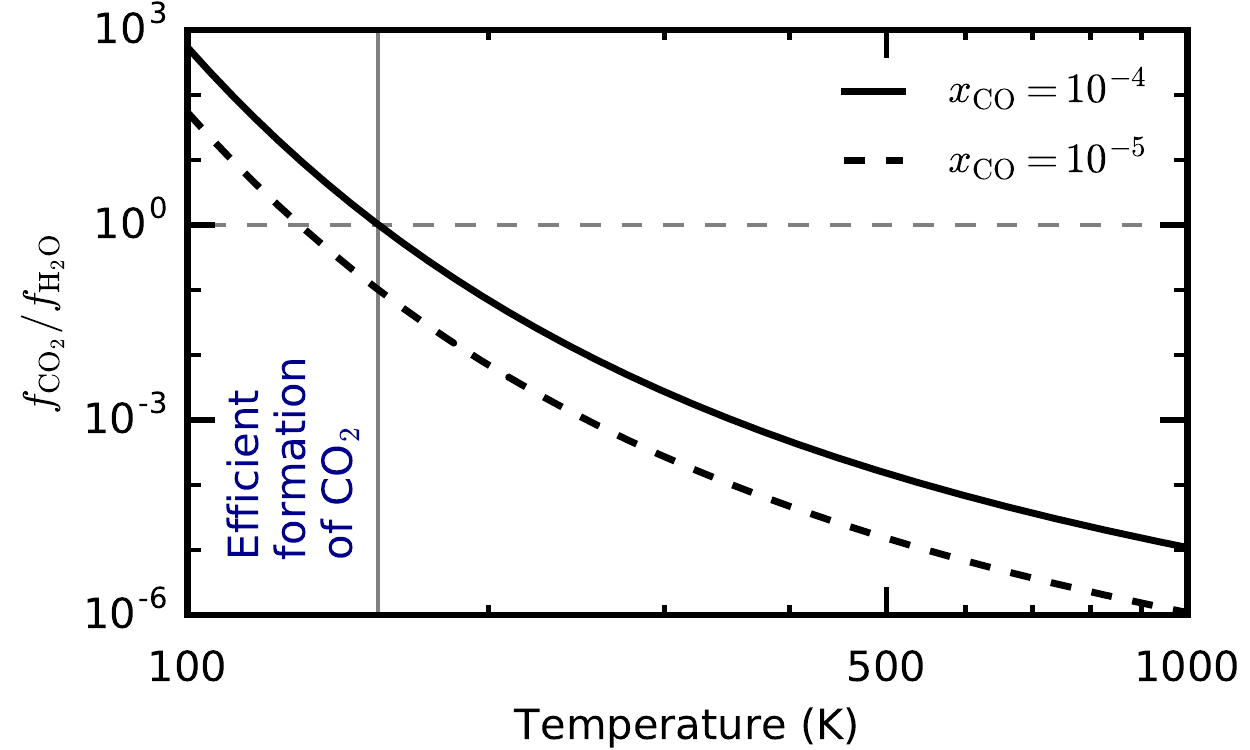}
\caption{\label{fig:ratefrac} Ratio of the \ce{CO2} to \ce{H2O} formation rate from a reaction of \ce{OH} with \ce{CO} and \ce{H2} respectively. Hydrogenation of \ce{OH} to \ce{H2O} dominates above 150 K. Formation times of \ce{CO2} and \ce{H2O} from \ce{OH} are faster than the inner disk mixing time for all temperatures considered here.}
\end{figure}

The \ce{CO2} formation reaction, \ce{CO + OH -> CO2 + H}, has competition from the \ce{H2O} formation reaction, \ce{OH + H2 -> H2O + H}. The hydrogenation of \ce{OH} has a higher activation energy, 1740 K, than the formation of \ce{CO2}, but since \ce{H2} is orders of magnitude more abundant than \ce{CO}, the formation of water will dominate over the formation of \ce{CO2} at high temperatures. The rate for \ce{CO2} formation is given by \citep{Smith2004}:
\begin{equation}
f_\mathrm{form, \ce{CO2}} =2.81\times 10^{-13} \, n_{\ce{CO}}n_{\ce{OH}} \exp{\left(-\frac{176 \mathrm{K}}{T}\right)}. 
\end{equation}
The formation rate for \ce{H2O} formation is given by \citep{Baulch1992}:
\begin{equation}
f_\mathrm{form, \ce{H2O}} = 2.05\times 10^{-12} \, n_{\ce{H2}} n_{\ce{OH}}  \left(\frac{T}{300 \mathrm{K}}\right)^{1.52} \exp{\left(-\frac{1740 \mathrm{K}}{T}\right)}. 
\end{equation}
This means that the expected $x_{\ce{CO2}}/x_{\ce{H2O}}$ fraction from formation is:
\begin{equation}
\frac{f_\mathrm{form, \ce{CO2}}}{f_\mathrm{form, \ce{H2O}}} = 0.14 \, x_{\ce{CO}} \left(\frac{T}{300 \mathrm{K}}\right)^{-1.52} \exp\left(\frac{1564 \mathrm{K}}{T}\right)
\end{equation}
This function is plotted in Fig.~\ref{fig:ratefrac} which shows that
the formation of \ce{CO2} is faster below temperatures of 150 K,
whereas above this temperature formation of water is faster. Above a
temperature of 300 K water formation is a thousand times faster than
\ce{CO2} formation. The implication is that gaseous \ce{CO2} formation
is only effective in a narrow temperature range, 50--150 K, and then
only if OH is present as well, requiring UV photons or X-rays to
liberate O and OH from CO or \ce{H2O}.

\subsubsection{Destruction of \ce{CO2}: Cosmic rays} 
\label{ssc:CRdestr}
Cosmic rays, 10--100 MeV protons and ions, have enough energy to penetrate deeply into the disk. Cosmic-rays can ionise \ce{H2} in regions where UV photons and X-rays cannot penetrate. The primary ionisation as well as the collisions of the resulting energetic electron with further \ce{H2} creates electronically excited \ce{H2} as well as excited \ce{H} atoms. These excited atoms and molecules radiatively decay, resulting in the emission of UV-photons \citep{Prasad1983}. These locally generated UV-photons can dissociate and ionise molecules. Here only the dissociation rate of \ce{CO2} is taken into account as ionisation of \ce{CO2} by this process is negligible. Following \cite{Heays2017} the destruction rate of species $X$ is written as:
\begin{equation}
k_X = \frac{\zeta_{\ce{H2}}x_{\ce{H2}}}{x_X} \int P(\lambda) p_X(\lambda) \mathrm{d}\lambda,
\end{equation}
where $\zeta_{\ce{H2}}$ is the direct cosmic ray ionisation rate of \ce{H2}, $x_X$ is the abundance of species $X$ w.r.t. \ce{H2}, $P(\lambda)$ is the photon emission probability per unit spectral density for which we use the spectrum from \cite{Gredel1987}. $p_X(\lambda)$ is the absorption probability of species $X$ for a photon of wavelength $\lambda$. This probability is given by:
\begin{equation}
p_X(\lambda) = \frac{x_X \sigma^{\mathrm{destr}}_X(\lambda)}{x_\mathrm{dust}\sigma^{\mathrm{abs}}_\mathrm{dust}(\lambda) + \sum_j x_j\sigma^\mathrm{abs}_j(\lambda)},
\end{equation}
where $\sigma^{\mathrm{i}}_X(\lambda)$ is the wavelength dependent destruction or absorption cross section of species $X$ and $x_\mathrm{dust}\sigma^{\mathrm{abs}}_\mathrm{dust}$ is the dust cross section per hydrogen molecule. 
For the calculation of the cosmic-ray dissociation rate of \ce{CO2} we assume that the destruction cross section in the UV is equal to the absorption cross section for \ce{CO2}, that is, every absorption of a photon with a wavelength shorter than 227 nm leads to the destruction of a \ce{CO2} molecule. 
For the calculations the cross sections from \cite{Heays2017} are used\footnote{UV cross-sections can be found here: \url{http://home.strw.leidenuniv.nl/~ewine/photo/}}. These cross sections can also be used to compute destruction rates for \ce{CO2} by stellar UV radiation in the surface layers of the disk. 

The dust absorption is an important factor in these calculations and can be the dominant contribution to the total absorption in parts of the spectrum. The dust absorption greatly depends on the dust opacities assumed. A standard ISM dust composition was taken following \cite{Weingartner2001}, the mass extinction coefficients are calculated using Mie theory with the \texttt{MIEX} code \citep{Wolf2004miex} and optical constants by \cite{Draine2003} for graphite and \cite{Weingartner2001} for silicates. Grain sizes are distributed assuming an MRN distribution starting at 5 nm, with varying maximum size are used. The resulting mass opacities and cross sections are shown in Fig.~\ref{fig:opacities}.

\begin{figure}
\centering
\includegraphics[width=\hsize]{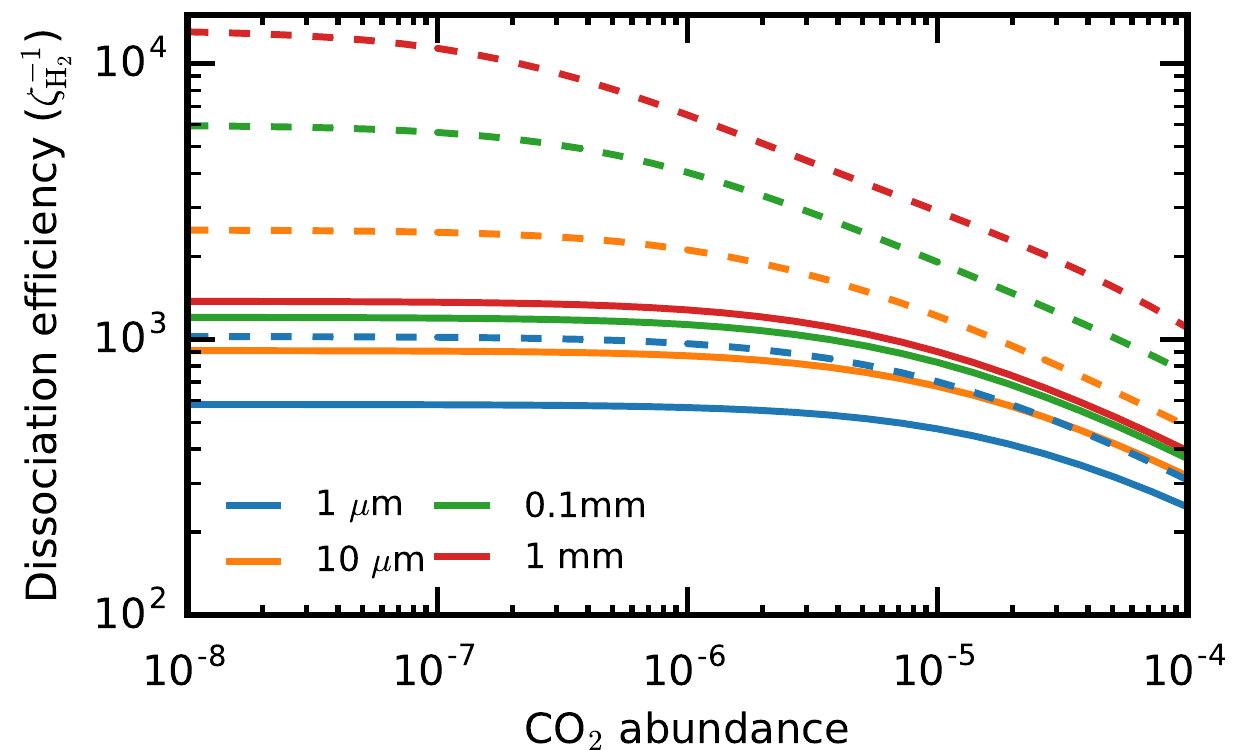}
\caption{\label{fig:CO2_destr} Dissociation rate of \ce{CO2} due to cosmic ray induced photons, for different dust size distributions. Solid lines are for condition inside of the \ce{H2O} iceline, dashed lines for conditions outside of the \ce{H2O} iceline. The efficiency multiplied by $\zeta_{
\ce{H2}}$ gives the \ce{CO2} dissociation rate. }
\end{figure}

Cosmic ray induced destruction rates for \ce{CO2} are calculated for each dust size distribution for a range of \ce{CO2} abundances between $10^{-8}$ and $10^{-4}$. The o-p ratio of \ce{H2}, important for the \ce{H2} emission spectrum, is assumed to be 3:1, representative for high temperature gas. The abundances of the other shielding species used in the calculation are shown in Table.~\ref{tab:shieldingspecies}. The destruction rate for \ce{CO2} is plotted in Fig.~\ref{fig:CO2_destr}.
	
\begin{table}
\centering
\caption{\label{tab:shieldingspecies}Gas-phase abundances assumed for the cosmic ray induced dissociation rate calculations}
\begin{tabular}{lll}
\hline\hline
Molecule & Abundance & Abundance\\
& inside \ce{H2O} iceline & outside \ce{H2O} iceline\\
\hline
\ce{H2} & 1  & 1\\
\ce{H} & $10^{-12}$   & $10^{-10}$\\ 
\ce{CO} & $10^{-5}$  &  $10^{-5}$\\
\ce{N2} & $10^{-5}$ & $10^{-5}$\\
\ce{CO2} & $10^{-8\cdots -4}$ & $10^{-8\cdots -4}$\\
\ce{H2O} & $10^{-4}$ & $10^{-8}$ \\
\hline
\end{tabular}

\end{table}
	
The \ce{CO2} destruction is faster for grown grains and low abundances
of \ce{CO2}. At abundances above $10^{-7}$, the strongest transitions
start to saturate, lowering the destruction rate with increasing
abundance. Even though the dust opacity changes by more than an order
of magnitude, the dissociation rates stay within a factor of 3 for all
\ce{CO2} abundances. This is due to the inclusion of \ce{H2O} into our
calculations. \ce{H2O} has large absorption cross sections in the same
wavelength regimes as \ce{CO2}. When \ce{H2O} is depleted, such as
would be expected between the \ce{H2O} and \ce{CO2} icelines, \ce{CO2}
destruction rates increase, especially for the largest grains. Even in
this optimal case, the destruction rate for \ce{CO2} is limited to
$2 \times 10^{-13}$ s$^{-1}$ for typical $\zeta_{\ce{H2}}$ of
$10^{-17}$ s$^{-1}$.

Aside from generating a UV field, cosmic-rays also create ions, the most important of these being \ce{He^+}. Due to the large ionisation potential of \ce{He}, electron transfer reactions with \ce{He^{+}} generally lead to dissociation of the newly created ion. For example, \ce{CO2 + He^{+}} preferably leads to the creation of \ce{O + CO^{+}}. Destruction of \ce{CO2} due to \ce{He^+} is limited by the creation of \ce{He^+} and the competition between \ce{CO2} and other reaction partners of \ce{He+}. The total reaction rate coefficient of \ce{CO2} with \ce{He+} is $k _\mathrm{ion, \ce{CO2}} = 1.14\times10^{-9}$~cm$^{3}$~s$^{-1}$\citep{Adams1976}. The three main competitors for reactions with \ce{He+} are \ce{H2O}, \ce{CO} and \ce{N2}, with reaction rate coefficients of $k _\mathrm{ion,\,\ce{H2O}} = 5.0\times 10^{-10}$, $k _\mathrm{ion,\,\ce{CO}} =1.6 \times 10^{-9}$ and $k _\mathrm{ion,\,\ce{N2}} =1.6 \times 10^{-9}$~cm$^{3}$~s$^{-1}$ respectively \citep{Mauclaire1978,Anicich1977,Adams1976}. Assuming a cosmic \ce{He+} ionisation rate of $0.65 \zeta_{\ce{H2}}$ \citep{Umebayashi2009}, the destruction rate for \ce{CO2} due to \ce{He+} reaction can be written as,
\begin{equation}
R_\mathrm{destr, \ce{He+}} = 0.65 \zeta_{\ce{H2}}\frac{x_{\ce{He}}}{x_{\ce{CO2}}} 
\frac{k _\mathrm{ion, \ce{CO2}}x_{\ce{CO2}}}{
\sum_X k _\mathrm{ion, X}x_{X}},
\end{equation}
where the sum is over all reactive collision partners of \ce{He^+} of which \ce{CO}, \ce{N2} and \ce{H2O} are most important.
For typical abundances of \ce{He} of 0.1 and \ce{N2}, \ce{CO} and \ce{H2O} of $10^{-4}$, $R_\mathrm{destr, \ce{He+}}$ is below $650 \zeta_{\mathrm{H}_2}$ for all \ce{CO2} abundances. Thus cosmic-ray induced photodissociation will always be more effective than destruction due to \ce{He+}.  

Altogether the destruction time-scale for midplane \ce{CO2} by cosmic
ray induced processes is long, $\sim 3$~Myr for a
$\zeta_{\ce{H2}} = 10^{-17}$~s$^{-1}$. The latter value is likely an
upper limit for the inner disk given the possibility of attenuation
and exclusion of cosmic rays \citep{Umebayashi1981,Cleeves2015}.

\subsubsection{Destruction of \ce{CO2}: Gas-phase reactions}
\label{ssc:Gasdestr}
In warm gas it is possible to destroy \ce{CO2} by endothermic reactions with \ce{H} or \ce{H2} \citep{TalbiHerbst2002}. The reaction 
\begin{equation}
\ce{CO2 + H -> CO + OH}
\end{equation} has an activation barrier of 13\,300 K and a pre-exponential factor of $2.5 \times 10^{-10}$ cm$^3$ s$^{-1}$ for temperatures between 300 and 2500~K \citep{TsangHampson1986} while the reaction 
\begin{equation}
\ce{CO2 + H2 -> CO + H2O}
\end{equation} has an activation barrier of 56\,900 K and has a pre-exponential factor of $3.3 \times 10^{-10}$ cm$^3$ s$^{-1}$ at 1000 K. This means that for gas at 300 K, an H$_2$ density of $10^{12}$ cm$^{-3}$, and a corresponding H density of 1 cm$^{-3}$, the rate for destruction by atomic hydrogen is $1.4 \times 10^{-29}$ s$^{-1}$, while the destruction rate by molecular hydrogen is $1.4 \times 10^{-80}$ s$^{-1}$. Both are far too low to be significant in the inner disk. However, if the atomic \ce{H} abundance is higher, destruction of \ce{CO2} by \ce{H} can become efficient at high temperatures. As such the destruction of \ce{CO2} becomes very dependent on the formation speed of \ce{H2} at high densities and temperatures. The \ce{CO2} abundance as a function of temperature for a gas-phase model is shown in Fig.~\ref{fig:CO2_temp}. Temperatures of $>$700 K are needed to lower the \ce{CO2} abundance below $10^{-7}$, even with a high atomic H abundance.

\subsection{Simulating spectra}
To compare the model abundances with observations, infrared spectra are
simulated with the thermochemical code DALI
\citep{Bruderer2012,Bruderer2013}. DALI is used to calculate a radial
temperature profile, from dust and gas surface density profiles and
stellar parameters. The viscous evolution model is initialised using
the same surface density distribution as the DALI model. The
temperature slope $q$ is taken so that the surface density slope $p$
is consistent with a self-similar solution $q = 1.5 - p = 0.9$. The
temperature at 1 AU is taken from the midplane temperatures as
calculated by the DALI continuum ray-tracing module. No explicit
chemistry is included in this version of DALI, but the abundances are
parametrised using the output of the dynamical model. The viscous model
run with $\alpha = 10^{-3}$, $u_f = 3$\,m\,s$^{-1}$ and a \ce{CO2}
destruction rate varying between $10^{-13}$ and
$10^{-9}$\,s$^{-1}$. The results from Sect.~\ref{ssc:Total_viscous}
show that these parameters represent the bulk of the model
results.

After simulating 1 Myr of evolution, the gas-phase \ce{CO2} abundance profile is extracted as function of temperature and interpolated onto the DALI midplane temperatures. The abundance is taken to be vertically constant up to the point where  $\mathrm{A}_\mathrm{V} = 1$. In some variations an abundance floor of $10^{-9}$ or $10^{-8}$ is used for cells with $\mathrm{A}_\mathrm{V} > 1$ to simulate local \ce{CO2} production. Using the resulting abundance structure, the non-LTE excitation of \ce{CO2} is calculated using the rate coefficients from \cite{Bosman2017}. Finally the DALI line ray-tracing module calculates the line fluxes for \ce{CO2} and its \ce{^{13}C} isotope.

For calculating the spectra, the same disk model is used as in \cite{Bruderer2015} and \cite{Bosman2017}. The model is based on the disk AS 205 N. The parameters for the DALI models can be found in Table~\ref{tab:modelsetup}. For more specifics on the modelling of the spectra and adopted parameters, see \cite{Bruderer2015} and \cite{Bosman2017}. To simulate the high gas-to-dust ratios that are inferred from water observations \citep{Meijerink2009}, the overall gas-to-dust ratio is set at 1000 throughout the disk. The artificially high gas-to-dust ratio does not affect any of the modelling, except for the line formation, which is only sensitive to the upper optically thin layers of the disk. 
High gas-to-dust ratios effectively mimic settled dust near the
midplane containing $\sim$90\% of the dust mass.

\begin{table}
\centering
\caption{\label{tab:modelsetup} Adopted standard model parameters for 
the DALI modelling.}
\begin{tabular}{l l c}
\hline
\hline
Parameter &  & Value\\
\hline
Star\\
Mass & $M_\star$ [$M_\odot$] & 1.0 \\
Luminosity & $L_\star$ [$L_\odot$] & 4.0 \\
Effective temperature & $T_\mathrm{eff}$ [K] & 4250 \\
Accretion luminosity & $L_\mathrm{accr}$ [$L_\odot$] & 3.3 \\
Accretion temperature & $T_\mathrm{accr}$ [K] & 10000 \\
\hline
Disk\\
Dustdisk mass & $M_\mathrm{dust}$ [$M_\odot$]& $2.9 \times 10^{-4}$ \\
Surface density index & $p$ &  0.9 \\
Characteristic radius&  $R_c$ [AU] & 46 \\
Inner radius&  $R_\mathrm{in}$ [AU] & 0.19 \\
Scale height index\tablefootmark{b} & $\psi$ & 0.11 \\
Scale height angle\tablefootmark{b} & $h_c$ [rad]& 0.18\\
\hline
DALI dust properties\tablefootmark{a}\\
Size & a[$\mu m$] & 0.005 -- 1000 \\
Size distribution & & $\mathrm{d}n/ \mathrm{d}a \propto a^{-3.5} $\\
Composition & & ISM \\
Gas-to-dust ratio & &1000 \\
\hline
Distance & d [pc] & 125 \\
Inclination & $i$ [$^{\circ}$] & 20\\
\hline
\vspace{-0.35cm}\\
\ce{^{12}CO2}:\ce{^{13}CO2} ratio & & 69\\
\hline
\end{tabular}
\tablefoot{
\tablefoottext{a}{Dust properties are the same as those used in \cite{Andrews2009} and \cite{Bruderer2015}. Dust composition and optical constants are taken from \cite{Draine1984} and \cite{Weingartner2001}.}
\tablefoottext{b}{Only used in the DALI model, not in the viscous disk model. The viscous disk model assumes a geometrically flat disk. }
}
\end{table}

\section{Results}
\label{sec:Results}
\subsection{Pure viscous evolution}
\label{ssc:PureViscous}
\begin{figure*}
\includegraphics[width=\hsize]{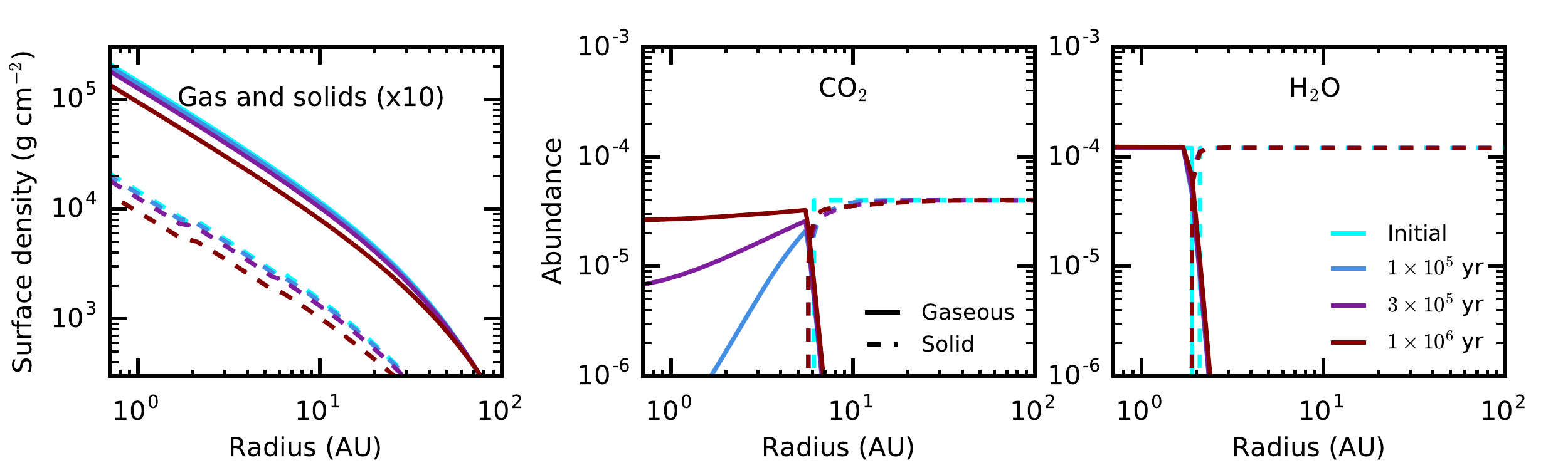}
\caption{\label{fig:nogrowthmodel} Time evolution series for a model without grain growth. This model assumes an $\alpha$ of 10$^{-3}$. \emph{Left}: Surface density of the gas (solid lines) and solids (dashed lines), the solid surface density is the sum of the dust and ice surface densities, it has been multiplied by a factor of 10 for visualization. \emph{Middle}: Abundance of \ce{CO2}. \emph{Right}: Abundance of \ce{H2O}. }
\end{figure*}

To start, a dynamical model without grain growth and without any chemistry
(except for freeze-out and desorption) is
investigated. Fig.~\ref{fig:nogrowthmodel} shows the time evolution
for a model with  $\alpha=10^{-3}$. As expected, the total gas and dust
surface densities barely change over $10^6$ yr. The gas and dust evolve viscously, some
mass is accreted onto the star while the outer disk spreads a little
bit. There are no changes in gas-to-dust ratio in the disk.

The water abundance in the right panel of Fig.~\ref{fig:nogrowthmodel} shows no evolution at all. This is to be expected because the only radial evolution is due to the gas viscosity, which affects gas and dust equally. Diffusion of the icy dust grains is equally quick as the diffusion of the water vapour. This means that all the water vapour that diffuses outwards is compensated by icy dust grains diffusing inwards.

The abundance of \ce{CO2} in the middle panel of Fig.~\ref{fig:nogrowthmodel} does show evolution. Initially there is only a little bit of \ce{CO2} gas in the inner disk, but a large amount of ice in the outer disk. The abundance gradient together with the viscous accretion makes the icy \ce{CO2} move inwards, filling up the inner disk with \ce{CO2} gas. This continues for about $1 \times 10^6$ yr until the gaseous abundance of \ce{CO2} is equal to the abundance of icy \ce{CO2}. This time-scale directly scales with the assumed $\alpha$ parameter. For $\alpha = 10^{-4}$ it takes more than 3 Myr to get a flat abundance profile in the inner disk. 

\subsection{Viscous evolution and grain growth}
\label{ssc:ViscousandGrowth}
\begin{figure*}
\includegraphics[width=\hsize]{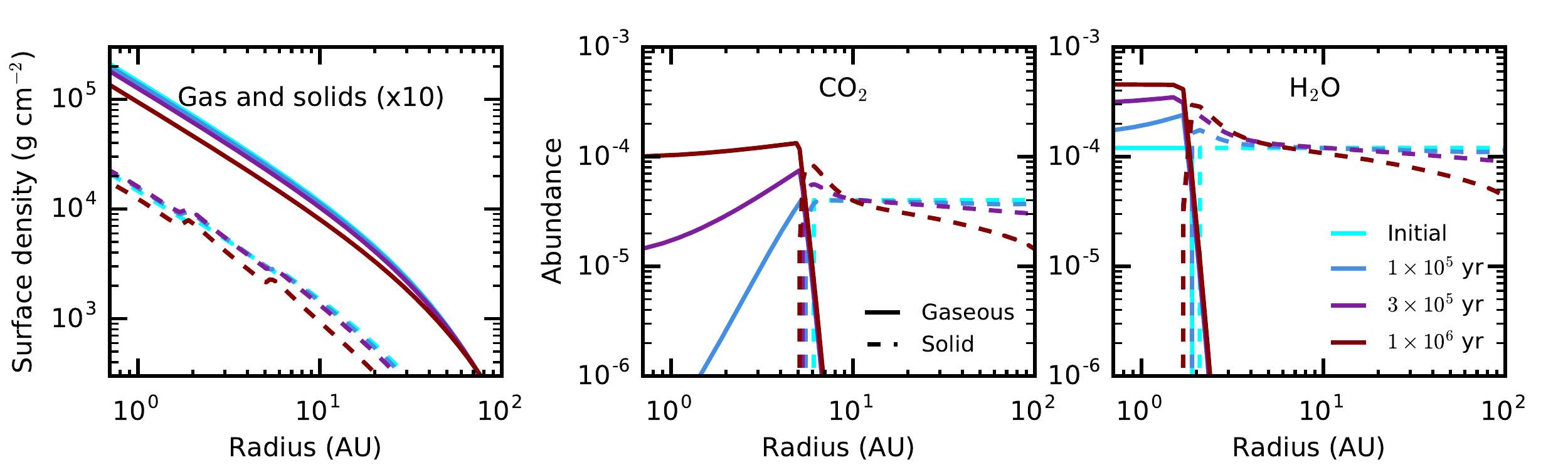}
\caption{\label{fig:growthmodel3ms} Time evolution series for a model with grain growth. This model assumes an $\alpha$ of 10$^{-3}$. The fragmentation velocity for this model is 3~m~s$^{-1}$. Panels as in Fig.~\ref{fig:nogrowthmodel}.}
\end{figure*}

\begin{figure*}
\includegraphics[width=\hsize]{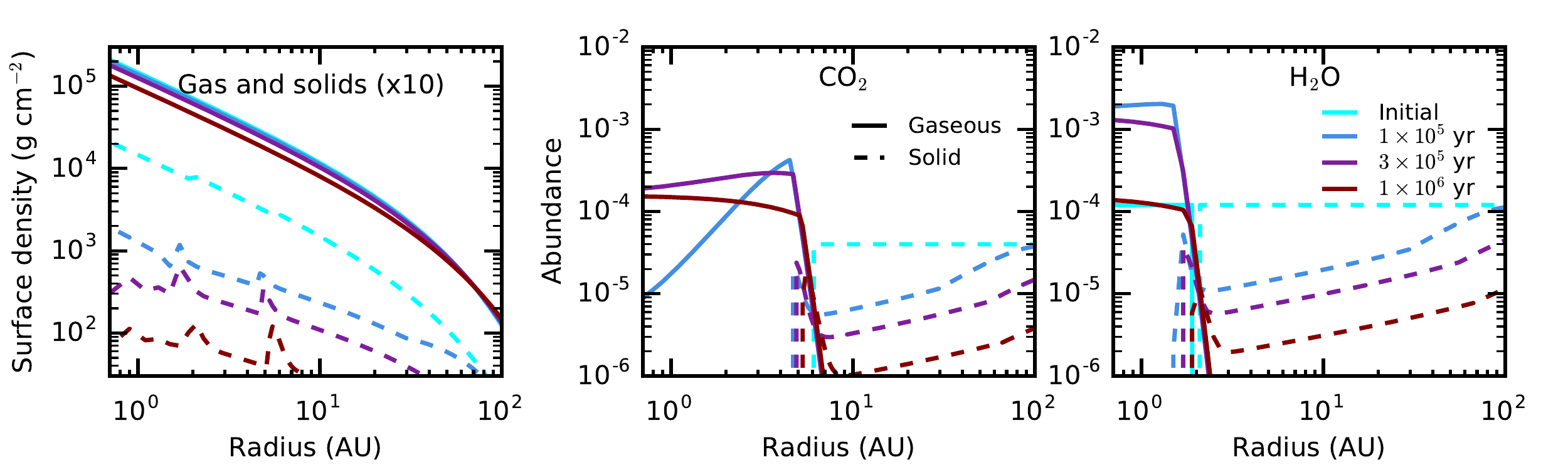}
\caption{\label{fig:growthmodel30ms} Time evolution series for a model with grain growth. The fragmentation velocity for this model is 30 m s$^{-1}$ and $\alpha = 10^{-3}$. The vertical scales are different from Figs.~\ref{fig:nogrowthmodel}~and~\ref{fig:growthmodel3ms}.  Panels as in Fig.~\ref{fig:nogrowthmodel}.}
\end{figure*}

For the models including grain growth, we tested a range of fragmentation velocities from 1 to 30 m s$^{-1}$. For some of these models, especially those with low fragmentation velocities and high $\alpha$, the results are indistinguishable from the case without grain growth. Figs.~\ref{fig:growthmodel3ms}~and~\ref{fig:growthmodel30ms} show two models where the effect of grain growth and resulting drift can be seen. 

Fig.~\ref{fig:growthmodel3ms} shows the surface density and abundance evolution for a model with a fragmentation velocity of 3 m s$^{-1}$ as appropriate for pure silicate grains. The surface density of the dust shows a small evolution due to radial drift. Instead of decreasing, the surface density of dust in the inner 4 AU slightly increases in the first 300'000 yr due to the supply of dust particles from outside this radius. 

The abundance profiles in Fig.~\ref{fig:growthmodel3ms} show distinct effects of radial drift. Both the gaseous \ce{H2O} and \ce{CO2} abundances are high at all times due to the influx of drifting icy grains. There is also a decrease in the abundance of ices at large radii. This is because the grains carrying the ices have moved inwards, increasing the gas-to-ice ratios. 

The models are not in steady state after 1 Myr. If these models are evolved further, at first the abundances of \ce{H2O} and \ce{CO2} will increase further. At some point in time the influx of dust will slow down, because a significant part of the dust in the outer disk will have drifted across the snowline. At this point the inner disk abundances of \ce{H2O} and \ce{CO2} start to decrease as these molecules are lost due to accretion onto the star but no longer replenished by dust from the outer disk. For the model shown in Fig.~\ref{fig:growthmodel3ms} the average gas-to-dust ratio in the disk would be around 500 after 3 Myr.

For higher $\alpha$ the effects of radial drift are limited as the high rate of mixing smooths out concentration gradients created by radial drift. At the same time the maximum grain size is limited due to larger velocity collisions at higher turbulence. As such, for $\alpha=0.01$ a fragmentation velocity higher than 10 m s$^{-1}$ is needed to see significant effects of radial drift and grain growth. 

For lower $\alpha$ the effects of grain growth get more pronounced, 90\% of the dust mass is accreted onto the star in 2.5 Myr for a fragmentation velocity of 1 m s$^{-1}$. In this time a lot of molecular material is released into the inner disk and peak abundances of $10^{-3}$ and $10^{-2}$ are reached for \ce{CO2} and \ce{H2O}, respectively, after 1 Myr of evolution. Due to the low turbulence in the gas, the volatiles released are not well mixed, neither inwards nor outwards, so there is no strong enhancement of the ice surface density just outside of the iceline. An overview of the different model evolutions can be found in App.~\ref{app:growthmodels}.

Increasing the fragmentation velocity allows grains in the model to grow to larger sizes, leading again to larger amounts of radial drift. Fig~\ref{fig:growthmodel30ms} shows the evolution of the same model as shown in Fig.~\ref{fig:growthmodel3ms}, but with a fragmentation velocity of 30~m~s$^{-1}$, representative for grains coated in water ice. The solid surface density distribution clearly shows the effects of grain growth. At all radii, mass is moved inwards at a high rate. When the silicate surface density has dropped by an order of magnitude, a local enhancement in the surface density of the total solids is seen at the icelines. This enhancement is about a factor of 2 for both icelines. 

The gas-phase abundances of \ce{CO2} and \ce{H2O} both show a large increase in the first $3 \times 10^{5}$ yr due to the large influx of icy pebbles. At later times, the abundances are decreased again as the volatiles are accreted onto the star. After $3 \times 10^6$ yr the \ce{H2O} and \ce{CO2} abundances are lower than $10^{-5}$.

\subsection{Viscous evolution and {\ce{CO2}} destruction}

The models shown in
Fig.~\ref{fig:nogrowthmodel}--\ref{fig:growthmodel30ms} without any
chemical processing predict \ce{CO2} abundances that are very high in
the inner regions, $10^{-5}$--$10^{-4}$, orders of magnitude higher
that the value of $\sim 10^{-8}$ inferred from observations
\citep{Pontoppidan2014,Bosman2017}. There are multiple explanations
for this disparity, both physical and chemical, which are discussed in \S 5. 
Here we investigate how large any missing chemical destruction route
for gaseous \ce{CO2} would need to be. As discussed in
Sect.~\ref{sec:chemproc} and App.~\ref{app:gasmodel}, midplane \ce{CO2}
can only be destroyed by cosmic-ray induced processes in the current
networks. However, both cosmic ray induced photodissociation and
\ce{He^+} production are an order of magnitude slower than the viscous
mixing time. We therefore introduce additional destruction of gaseous
\ce{CO2} with a rate that is a constant over the entire disk. This
rate is varied between $10^{-13}$ and $10^{-9}$~s$^{-1}$ to obtain
agreement with observations. For comparison, the cosmic ray induced
process generally have a rate of the order of $10^{-14}$~s$^{-1}$ (see~Sect.~\ref{ssc:CRdestr}). The rate is implemented as an effective destruction rate, so
there is no route back to \ce{CO2} after destruction. To get the
absolute rate one has to also take into account the reformation
efficiency, but that can be strongly dependent on the destruction
pathway, of which we are agnostic. These efficiencies are discussed in
\S 5 and App.~\ref{app:chem}.

\begin{figure*}
\includegraphics[width=\hsize]{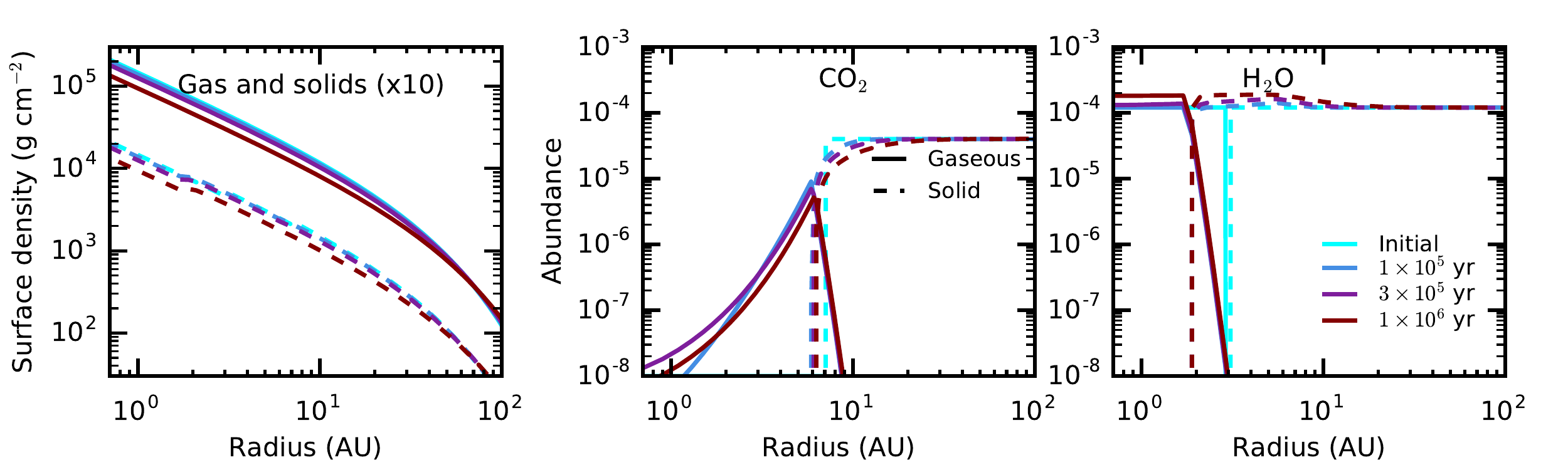}
\caption{\label{fig:nogrowth_flat} Time evolution series for a model without grain growth but with a constant destruction rate of gaseous \ce{CO2} of $10^{-11}$ s$^{-1}$. This model assumes an $\alpha$ of 10$^{-3}$. Panels as in Fig.~\ref{fig:nogrowthmodel}, but with different vertical scales.}
\end{figure*}

Fig.~\ref{fig:nogrowth_flat} shows the abundance evolution for a model
where only gaseous \ce{CO2} is destroyed, at a rate of
$10^{-11}$~s$^{-1}$. In this case the innermost parts of the disk are
empty of \ce{CO2}, whereas the abundance of \ce{CO2} reaches values
close to the initial ice abundances around the iceline. Due to the
constant destruction of \ce{CO2} near the iceline, the actual ice
abundance of \ce{CO2} is also lower than the initial value.

The left panel of Fig.~\ref{fig:destr_rate_abu} shows the \ce{CO2} abundance distribution for models with $\alpha = 10^{-3}$. The abundance profiles after 1 Myr of evolution are presented, when a semi-steady state has been reached.
The peak abundance and peak width of the \ce{CO2} gas abundance
profile both depend on the assumed destruction rate: a higher
rate leads to a lower peak abundance and a narrower peak, while a
lower rate leads to the opposite. A rate of $\sim 10^{-11}$~s$^{-1}$ or higher is needed to decrease the average gaseous \ce{CO2} abundance below the observational limit (see below). Increasing the rate moves the
location of the iceline further out, as the total available \ce{CO2} near the iceline decreases. The viscosity also influences the
width of the abundance profile, higher viscosities lead to a broader
abundance peak.

\subsection{Viscous evolution, grain growth and {\ce{CO2}} destruction}
\label{ssc:Total_viscous}
The next step is to include \ce{CO2} destruction in viscous evolution
models with grain growth. Here only models that retain the disk dust
mass, that is, models that have an overall gas-to-dust ratio smaller
than 1000 after 1 Myr of evolution are considered since there is no
observational evidence for very high gas-to-dust ratios. This means that the grains in our models do not reach pebble sizes such as have been inferred from observations \citep[e.g. ][]{Perez2012,Tazzari2016}. It is unclear how these large grains are retained in the disk as the radial drift timescale for these particles is expected to be shorter than the disk lifetime \citep[e.g. ][]{Birnstiel2010,Krijt2016Panoptic}. 
The requirement of dust retention restricts our models to $\alpha = 10^{-4}$ and
$u_f = 1$~m~s$^{-1}$, $\alpha = 10^{-3}$ and
$u_f = 1,\,3$~m~s$^{-1}$, and $\alpha = 10^{-2}$ and
$u_f = 1,\,3,\,10$~m~s$^{-1}$.

\begin{figure*}
\centering\includegraphics[type=pdf,ext=.pdf,read=.pdf,width = \hsize]{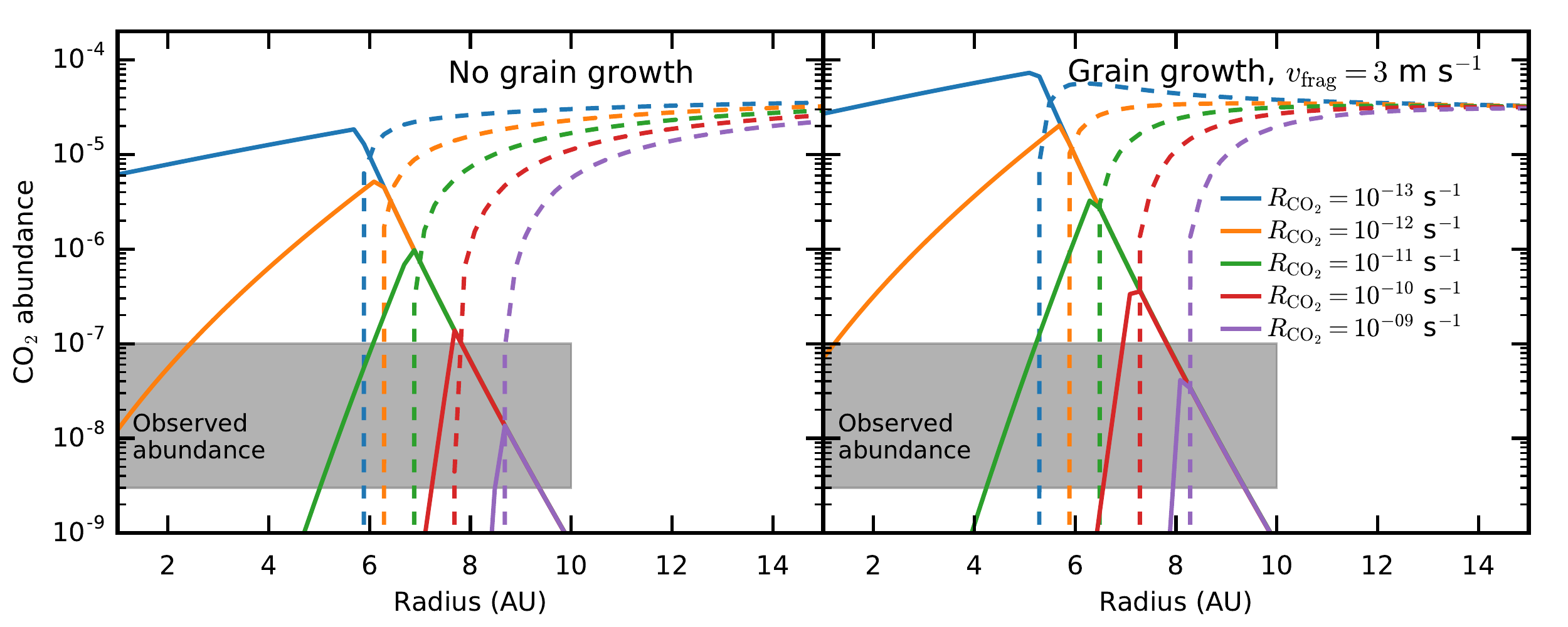}
\caption{\label{fig:destr_rate_abu} Abundance of \ce{CO2} as function of radii for models with different destruction rates of \ce{CO2}. These model use 
$\alpha = 10^{-3}$ and $u_f = 0$~m~s$^{-1}$ (\emph{left}) and $u_f = 3$~m~s$^{-1}$ (\emph{right}) . Abundance profiles after 1 Myr of evolution are shown. At this point the disk has reached a semi steady state. A destruction rate of at least $10^{-11}$~s$^{-1}$ is needed to keep the averaged \ce{CO2} abundance below the observational limit. The iceline is further out for models with higher destruction rates as the destruction of gas-phase \ce{CO2} lowers the total abundance of \ce{CO2} within 10 AU where \ce{CO2} can sublimate. } 
\end{figure*}

The abundance profiles for different \ce{CO2} destruction rates are shown in the right panel of Fig.~\ref{fig:destr_rate_abu} for the model with $\alpha = 10^{-3}$ and $u_f =\ 3$~m~s$^{-1}$. The abundance profiles are determined after 1 Myr, when the disk is in a semi-steady state. A destruction rate higher  than $10^{-12}$~s$^{-1}$  will create an abundance profile with clear abundance maximum at the iceline, but for this rate $10^{-12}$~s$^{-1}$ the disk surface area averaged \ce{CO2} abundance is still higher than the observed value for both cases. Only for a rate of around  $10^{-10}$ s$^{-1}$ do the models predict an abundance profile with a disk surface area averaged \ce{CO2} abundance that is consistent with the high end of the observations. Only the models with destruction rate higher than $10^{-9}$~s$^{-1}$ have a peak in the abundance that is consistent with the maximal inferred abundances from the observations. As shown in Fig.~\ref{fig:destr_rate_abu} the model with a rate of $10^{-11}$ s$^{-1}$ will create a spectrum consistent with the observations, even though the average \ce{CO2} abundance in higher than the observed abundance limit.

For the selected models, the effect of different fragmentation velocities is small, with differences in peak abundances less than a factor of five between similar models, as seen in Fig.~\ref{fig:destr_rate_abu}.\ This is unsurprising as the selection criteria for the models only consider models that have a transport velocity of solids that is less than an order of magnitude faster than the transport velocity of the model without grain growth and radial drift. For models with faster growth it becomes more arbitrary to give a representative abundance profile as semi steady state is not reached at any point in time before the disk is depleted of most of the dust and volatiles in the inner regions. However, for very young disks that may have very efficient radial drift of grains, a high gas-phase \ce{CO2} abundance is expected within the \ce{CO2} iceline unless the destruction rate is $10^{-11}$ s$^{-1}$ or higher.

\subsection{Model spectra}
\label{ssc:Dalimodels}
The spectral modelling the focuses on the \ce{CO2} bending mode around centred at 15$\mu$m. This bending mode has a strong $Q-$branch that has been observed by \textit{Spitzer} in protoplanetary disks \citep{Carr2008,Pontoppidan2010}. This region also has a weaker feature due to the $Q-$branch of \ce{^{13}CO2} at 15.42 $\mu$m which can be used, together with the \ce{^{12}CO2} $Q-$branch, to infer information on the abundance structure \citep{Bosman2017}.

\begin{figure}
\centering
\includegraphics[type=pdf,ext=.pdf,read=.pdf,width = \hsize]{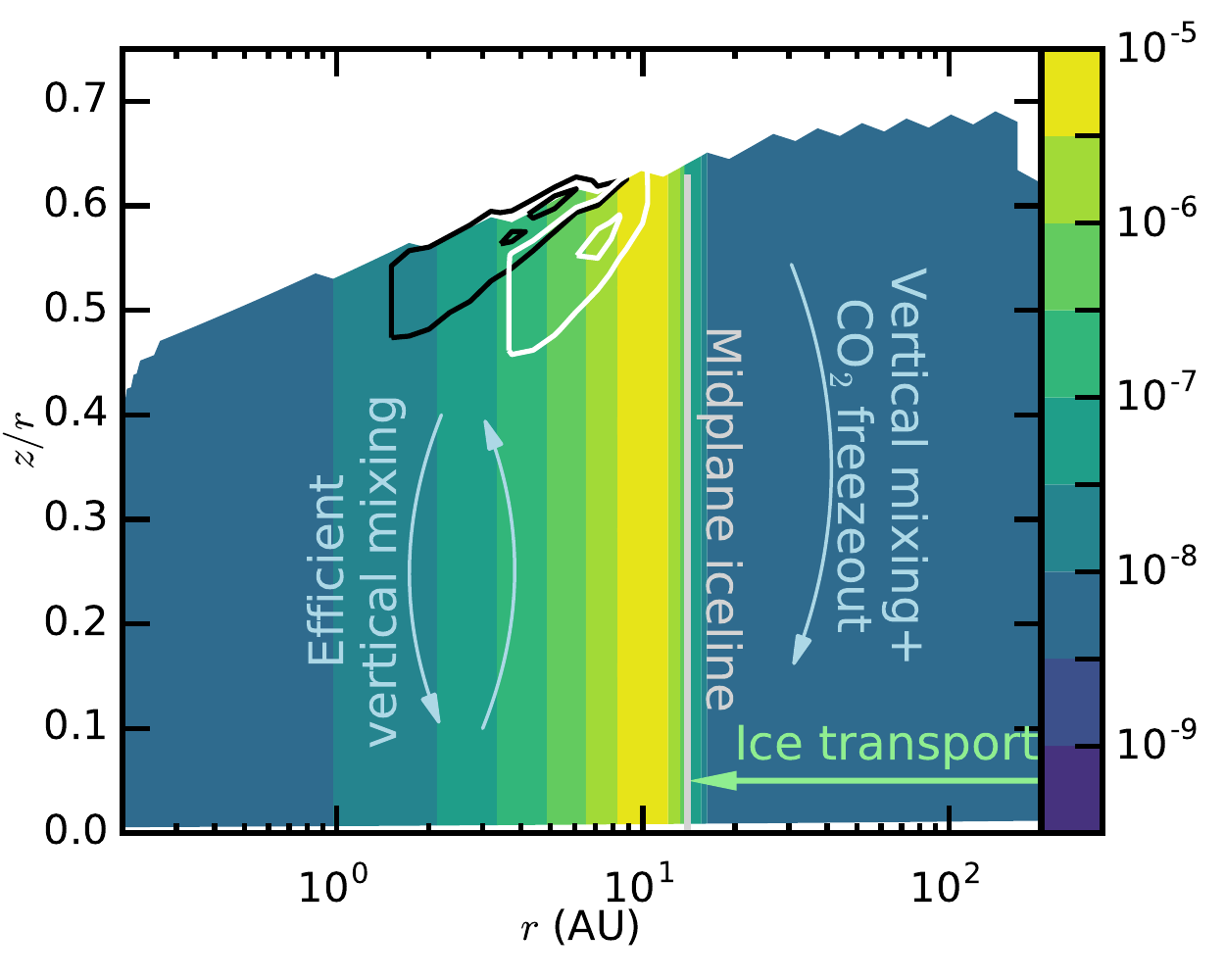}
\caption{\label{fig:DALI_abu_line} 2D \ce{CO2} abundance distribution from the $R_{\mathrm{destr, }\ce{CO2}} = 10^{-11}$ model with an abundance floor of $10^{-9}$. Black and white lines show the regions where 25\% and 75\% of the emission of respectively \ce{^{12}CO2} and \ce{^{13}CO2} is emitted from. The \ce{^{13}CO2} emission is more concentrated around the region with the abundance enhancement.}
\end{figure}

For the model with $\alpha = 10^{-3}$ and $R_{\mathrm{destr,\,}\ce{CO2}} = 10^{-11}$, the 2D abundance distribution assumed for the ray-tracing is shown in Fig.~\ref{fig:DALI_abu_line}. The enhanced \ce{CO2} abundance around the iceline shows up clearly. The results of the continuum ray-tracing can be found in Fig.~5 of \cite{Bosman2017}. The temperature at 1 AU is 320 K. This is higher than the value assumed for the models in the previous sections and is in line with the high luminosity of the modelled source of $7.7\,\mathrm{L}_\odot$. The higher temperature moves the \ce{CO2} iceline slightly further out. The abundance profiles from the viscous model with dust growth and \ce{CO2} destruction are still very similar to those from the previous section. 

Fig.~\ref{fig:DALI_abu_line} also shows the emitting region of the \ce{CO2} $Q(6)$ line for both isotopologues. This is one of the strongest lines in the \ce{CO2} spectrum and part of the $Q$-branch. As such it is a good representative for the complete $Q$-branch.
Due to its high abundance, the \ce{^{12}CO2} lines have a high optical depth over a larger area in the disk. As such a large part of the disk contributes to the emission of the \ce{^{12}CO2} lines. For \ce{^{13}CO2} the emission is from a more compact area, close to the peak in \ce{CO2} abundance. 

\begin{figure}
\centering
\includegraphics[type=pdf,ext=.pdf,read=.pdf,width = \hsize]{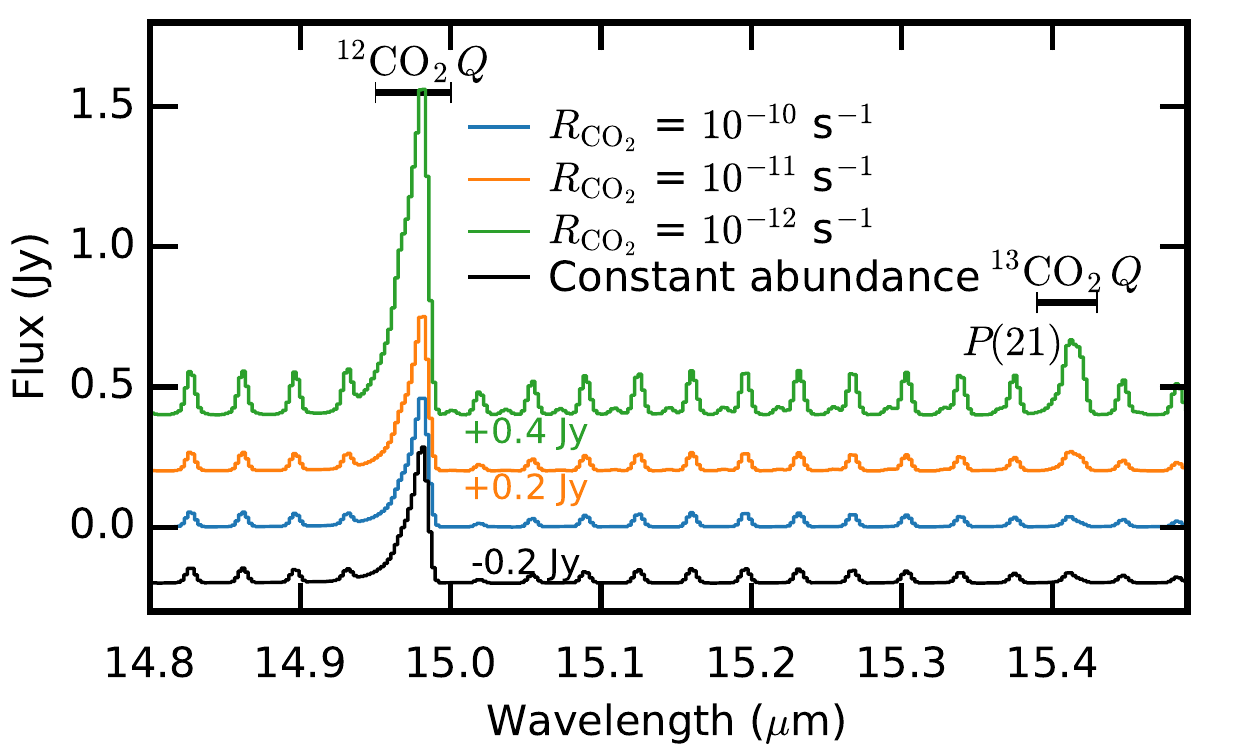}
\caption{\label{fig:DALI_spectrum-8} Spectra \ce{^{12}CO2} and \ce{^{13}CO2} for models with different destruction rates at a spectral resolving power $R=2200$ appropriate for JWST-MIRI. The abundance floor is taken to be $10^{-8}$. The black line shows the model results for a constant abundance as used in \cite{Bosman2017}. The regions were the $Q$-branch flux is extracted is shown in the black lines. The \ce{^{13}CO2} $Q$-branch is blended with the \ce{^12CO2} $P(23)$ line. The neighbouring $P(21)$ line is indicated. Note that the models with destruction rates of $10^{-11}$ and $10^{-10}$ s$^{-1}$ are indistinguishable in their \ce{^{12}CO2} spectra.}
\end{figure}

Fig.~\ref{fig:DALI_spectrum-8} shows \ce{CO2} spectra for different
destruction rates and abundance floor of $10^{-8}$.  Line-to-continuum
ratios for the individual $R$ and $P$ branch lines are all greater
than 0.01 and are detectable by \textit{JWST} \citep{Bosman2017}.
Note that only even $J$ levels exist in the vibrational ground state,
so $P$, $Q$ and $R$ branch lines exist only every other $J$ for this
band.  Both the \ce{^{12}CO2} and the \ce{^{13}CO2} $Q$-branch fluxes
are influenced by the different destruction rates. For a rate of
$10^{-12}$~s$^{-1}$ the \ce{^{12}CO2} $Q$-branch is a factor of two
brighter than for the higher rates, which give fluxes very similar in
magnitude for the \ce{^{12}CO2} $Q$-branch. For \ce{^{13}CO2} the
model with a destruction rate of $10^{-12}$~s$^{-1}$ is a least a
factor of five brighter than the models with higher rates of
$10^{-11}$~s$^{-1}$ and $10^{-10}$~s$^{-1}$. Since the latter two \ce{{^12}CO2} spectra are indistinguishable observationally, a conservative lower limit to the destruction rate of $10^{-11}$ s$^{-1}$ is chosen. For \ce{^{13}CO2}
however, there is a factor of three difference between the two higher
rate models; here the \ce{^{13}CO2} $Q$-branch flux is determined by
subtracting the contribution from the \ce{^{12}CO2} $P(23)$-line
assuming it is the same as that of the neighbouring $P(21)$ line.

\begin{figure}
\centering
\includegraphics[type=pdf,ext=.pdf,read=.pdf,width = \hsize]{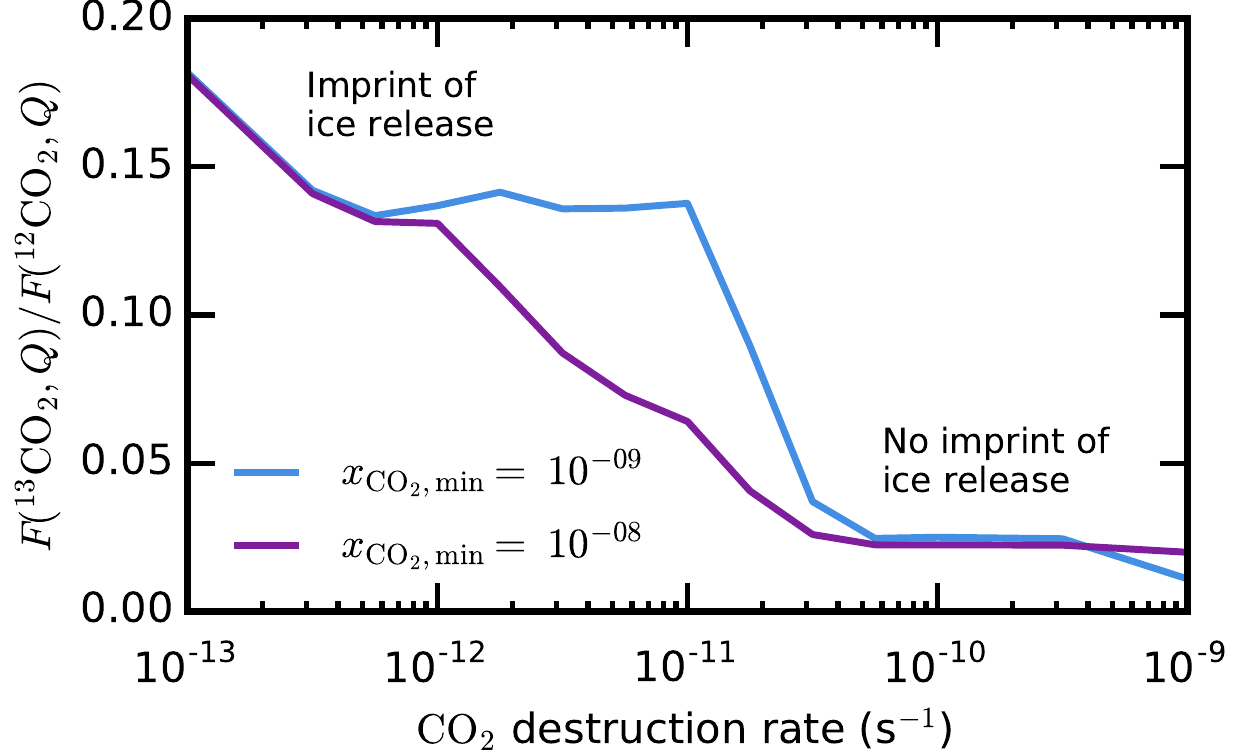}

\caption{\label{fig:DALI_fluxratios} Flux ratios of the \ce{^{13}CO2} and \ce{^{12}CO2} $Q$-branches,
determined by integrating the region shown in Fig.~\ref{fig:DALI_spectrum-8}. The \ce{^13CO2} $Q$-branch flux is corrected for the emission of the $P(23)$ line by subtraction of the neighbouring $P(21)$ line flux.}
\end{figure}

The flux ratio of the \ce{^{13}CO2} $Q-$branch over the \ce{^{12}CO2}
$Q$-branch is shown in Fig.~\ref{fig:DALI_fluxratios} for a range of
\ce{CO2} destruction rates. 
Three
regimes can be seen. For ratios larger than $\sim 0.13$ the \ce{CO2}
released from the ice dominates the abundance profile and spectrum,
due to the relatively low destruction rate. The high \ce{CO2} column,
due to the high abundance around the iceline sets the small flux
ratio. At ratios smaller than $~0.03$, corresponding to destruction
rates larger than $5\times 10^{-11}$ s$^{-1}$, the \ce{CO2} released from the
ice at the iceline is destroyed so fast that it can not leave an
imprint on the $Q$-branch fluxes. In the region in between, both the
\ce{CO2} released at the iceline as well as the upper layer \ce{CO2}
influence the spectra, as such the flux ratio not only depends on the
\ce{CO2} destruction rate, but also on the \ce{CO2} abundance floor
used.

Note that \cite{Bosman2017} show that a high \ce{^{13}CO2}
$Q-$branch over \ce{^{12}CO2} $Q$-branch flux ratio 
may also be an indication of a high abundance in the inner
disk. More generally it can be said that 
a high flux ratio indicates that the \ce{CO2} responsible for the
emission is highly optically thick. In these cases the width of the
\ce{^{13}CO2} $Q-$branch can be a measure for the temperature of this
emitting gas, which can be related to the radial location of the
origin of the emission. An analysis like this will probably only be
possible with disk specific modelling.

\subsubsection{Modelling uncertainties}

The spectral modelling done here is meant as an illustration to what
could be possible with \textit{JWST}-MIRI observations and to show
what kind of spectral features could hold clues to new insights into
disk dynamics. As such the model setup is not entirely self
consistent. Foremost, the temperature profiles calculated from the
radiative transfer are not exactly the same as the temperature profile
used in the viscous disk model. Furthermore, the starting point for
our viscous evolution is rather arbitrary.

Another uncertainty is the vertical
distribution of the \ce{CO2} sublimated from the ices. The assumed
turbulent viscosity would not only mix material radially, but also
vertically, and the \ce{CO2} that comes off the grain near the midplane
should make it up to the upper layers of the disk where the infrared
emission is generated. 

The vertical dispersion of a species injected into the disk midplane grows with time as $\sigma_\mathrm{X}=\sqrt{D_\mathrm{gas}t}$ \citep{Ciesla2010}. A steady state is reached if $\sigma_X = H$, the scale height of the disk, at which point a constant abundance is reached. As such we can estimate the time 
needed for vertical mixing. For a self-similar disk, this time scale becomes:
\begin{equation}
\label{eq:vertmixing}
t_\mathrm{vert,\ eq} = 9.1 \left(\frac{0.01}{\alpha}\right) \left(\frac{r }{1 \mathrm{AU}}\right)^{0.5 + q} \left(\frac{200}{T_\mathrm{1 AU}}\right)^q \left(\frac{\mathrm{h_{\mathrm{FWHM}}}}{0.05}\right) \mathrm {yr}.
\end{equation}
If a similar mixing speed ($\alpha$) is assumed for the
radial and vertical processes, then vertical mixing should happen well
within the lifetime of the disk for locations near the \ce{CO2}
iceline.

Given these uncertainties, a simple constant abundance up to a certain
height has been chosen. However higher up in the disk,
photodissociation by stellar radiation and processing by stellar
X-rays become important. \ce{CO2} is removed in our model in the
harshest environments ($A_V<1$ mag), but \ce{CO2} photodissociation is still
possible in the upper disk regions where \ce{CO2} is included.  The
location and thickness of this layer is discussed in
Sect.~\ref{sss:UV}~and~\ref{sss:X-ray}

\section{Discussion}
\label{sec:Discussion}

\begin{figure*}
\centering
\includegraphics[width=\hsize]{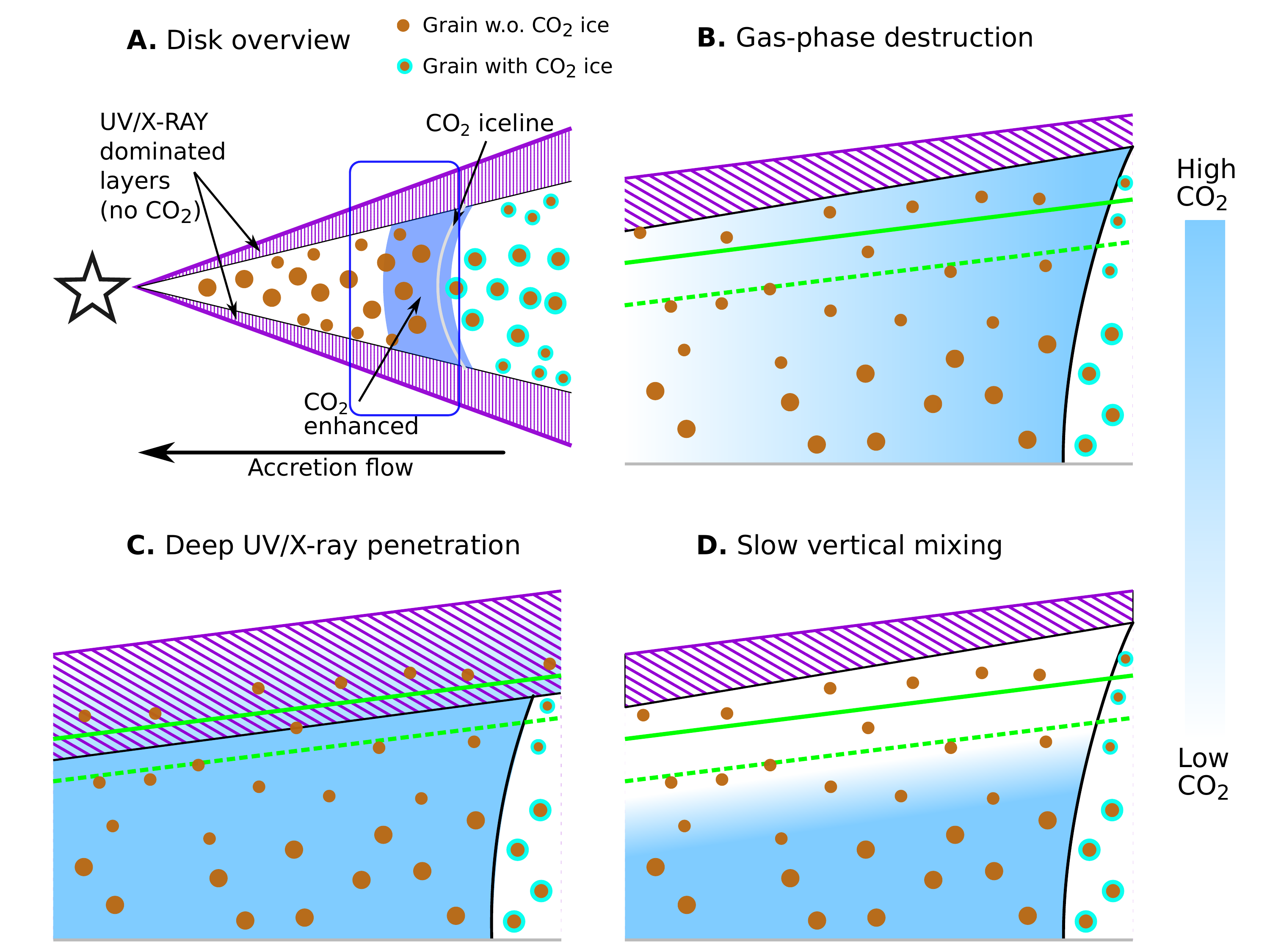}
\caption{\label{fig:schematic} Schematic representation of the \ce{CO2} abundance near the iceline in the $r-z$ plane. Panels B, C and D approximately represent the top half of the blue box in panel A. The straight green lines denote the $\tau = 1$ surface for \ce{^{12}CO2} (solid) and \ce{^13CO2} (dashed). Panels B, C and D show the abundance structure under different mechanism to explain the difference in \ce{CO2} abundance between the predictions of the viscous model and the IR observations. Panel B represents the case for a constant gas-phase destruction rate due to some as yet unidentified process (Sect.~\ref{ssc:Total_viscous}). Panel C shows the abundance structure for deep UV/X-ray penetration (Sect.~\ref{sss:UV}~and~\ref{sss:X-ray}). In more extreme cases, the \ce{^{13}CO2} $\tau = 1$ surface can also be in the UV/X-ray dominated region. Panel D shows the abundance structure for negligible vertical mixing (Sect.~\ref{sss:vmixing}). Image not to scale. }
\end{figure*}

Viscous disk models including only freeze-out and desorption of
\ce{CO2} predict a higher \ce{CO2} abundance than observed. This
conclusion holds even for models without grain growth and radial
drift. The discussion is divided into two sections, first the gas and
grain surface chemistry is discussed to explore alternative chemical
destruction routes. Subsequently physical effects are investigated,
some of which also imply chemical effects to happen.
Fig~\ref{fig:schematic} shows the \ce{CO2} abundance structure in the
disk with some of the mechanisms discussed here that could obtain agreement
between observations and models.

\subsection{Chemical processes}
\label{ssc:Chemdisc}
In the scenario that the \ce{CO2} released at the iceline is
destroyed, an effective destruction rate of $10^{-11}$ s$^{-1}$ or
higher is required to match the observed and computed abundances.  In
Sect.~\ref{ssc:CRdestr}~and~Sect.~\ref{ssc:Gasdestr} a number of known
destruction processes for midplane \ce{CO2} were discussed, but none of these
chemical processes is able to account for the destruction rate of
$10^{-11}$ s$^{-1}$. 

\subsubsection{Clues from high-mass protostars?}

Our finding that a \ce{CO2} ice desorption model is inconsistent with
\ce{CO2} gas-phase observations is not the first to do so. For high
mass protostars \cite{Boonman2003CO2} noted a similar disagreement
between simple in-fall and desorption models, which predict \ce{CO2}
gas-phase abundances around $10^{-5}$ whereas observations are more
consistent with an abundance of $\sim 10^{-7}$, a two orders of
magnitude difference. Assuming that the \ce{CO2} ice sublimates and is
destroyed within the dynamical time of $10^{4}$ yr, a destruction rate
of at least $10^{-11}$ s$^{-1}$ is needed for protostars as well, of
the same order as the rate needed in disks. If the \ce{CO2}
destruction mechanism is similar in both types of sources, then this
would most likely be a chemical destruction mechanism with no or a
weak dependence on total density since densities in protostellar
envelopes are orders of magnitude lower than in the inner disk
midplanes and the physical environments are very different. However,
as argued below the existence of such a chemical pathway is unlikely,
except for destruction by UV photons or X-rays.

\subsubsection{Alternative gas phase destruction routes}

Here we investigate whether a chemical pathway could be missing in the
chemical networks. Since ion-molecule reactions initiated by He$^+$
have already been discarded, the assumption is that this destruction
comes from a neutral-neutral reactive collision. The rate for a
two-body reaction can be written as:
\begin{equation}
R_\mathrm{\ce{CO2}, destr} = f_X k_\mathrm{2-body} n_\mathrm{gas} x_X
\end{equation}
where $n_\mathrm{gas}$ is the gas number density, $k_{\rm 2-body}$ is
the two-body reaction rate coefficient, $x_X$ is the abundance of the
collision partner and $f_X$ is the branching ratio to any product that
does not cycle back to \ce{CO2}. Near the iceline densities are around
$10^{13}$~cm$^{-3}$, a typical rate coefficient for a barrier-less
neutral-neutral reaction is around $10^{-11}$~cm$^{3}$ s$^{-1}$. As
such the partner abundance needs to be higher than $10^{-13}$ to
destroy \ce{CO2} at a fast enough rate. The only species that have a
nearly barrier-less reaction with \ce{CO2} in either the UMIST
\citep{McElroy2013} or KIDA \citep{Wakelam2012} databases are \ce{C},
\ce{CH2} and \ce{Si}. Reaction rates for \ce{CO2} with \ce{C} to form
two \ce{CO} molecules and with \ce{CH2} to form \ce{CO + H2CO} are
$10^{-15}$~cm$^{3}$~s$^{-1}$ and
$3.9\times 10^{-14}$~cm$^{3}$~s$^{-1}$ \citep{TsangHampson1986},
however, both rates are high temperature estimates and extrapolated to
low temperatures \citep{Hebrard2009}. The rate for \ce{CO2 + Si -> CO
  + SiO} has been measured at room temperature to be
$1.1\times 10^{-11}$~cm$^{3}$~s$^{-1}$ \citep{Husain1978},
\cite{McElroy2013} assume a small energy barrier of 282 K, in line
with the high temperature measurements. Other species in the databases
that have an exothermic reaction with \ce{CO2} are \ce{CH} and
\ce{N}; both of these reactions are thought to have a barrier of 3000
and 1710 K respectively \citep{Mitchell1984,Avramenko1967}. Other
neutral-neutral reactions in the chemical networks are highly
endothermic, for example \ce{CO2 + S -> CO + SO} is endothermic by
$\sim20000$ K \citep{Singleton1988}, effectively nullifying this
reaction in the temperature range we are interested in.

\ce{C}, \ce{CH2} and \ce{Si} are, due to their high reactivity, only
present in very low quantities in the gas-phase with abundances at the solver accuracy limit of $10^{-17}$ for \ce{C} and \ce{CH2}, and $10^{-16}$ for \ce{Si} and thus can hardly account
for the destruction of a significant amount of \ce{CO2} unless it is possible to quickly reform the initial reactant from the reaction products. The reactions products, \ce{CO}, \ce{H2CO} and \ce{SiO}, are again very
stable, thus it is unlikely that these molecules quickly react further
to form the initial reactants again. This means that the abundance of \ce{C}, \ce{CH2} and \ce{Si} will go down with time if more and more \ce{CO2} is added.

\subsubsection{Alternative ice destruction routes}

The low \ce{CO2} abundance observed could also be explained if
\ce{CO2} is converted to other species while in the
ice. \cite{Bisschop2007} report that \ce{CO2} bombarded with \ce{H}
does not lead to HCOOH at detectable levels in their experiments.
There are no data available on grain surfaces reactions of
\ce{CO2} ice with \ce{N}, \ce{Si} or \ce{C}. If any of these reaction
pathways were to destroy \ce{CO2} in the ice, they should be very
efficient, $\sim 99\%$ conversion, but at the same time be able to
explain the high \ce{CO2} ice abundances in comets. This quickly
limits the temperature range in which these reactions can be efficient
to 30--80 K.

In short, due to the high stability of \ce{CO2}, only a handful of
highly reactive radicals or atoms can destroy \ce{CO2}, whether in the
gas or ice. None of these species is predicted to be abundant enough
to quickly destroy the large flux of \ce{CO2} that is brought into the
inner disk due to dynamic processes of CO$_2$-containing icy grains.

\subsection{Physical processes}
\label{ssc:Discusphys}
The abundance profiles computed in this work all assume that the
\ce{CO2} can move freely through the disk, both in radial and vertical
directions, while being shielded from UV and X-ray radiation. It is
also assumed that all \ce{CO2} ice sublimates at the \ce{CO2} iceline
and that \ce{CO2} is highly abundant in the ice, as found in the
ISM. Several physical processes could make these assumptions invalid.
An important constraint is that due to the difference between the ice
abundances (around $10^{-5}$) and the inferred gas abundance
($<10^{-7}$) any process needs reduce the \ce{CO2} abundance in the
infrared emitting layers by at least two orders of magnitude.

\subsubsection{Impact of reduced turbulence}
\label{sss:vmixing}
The current models assume a standard viscously spreading and evolving
disk with a constant $\alpha$ value.  Magneto-hydrodynamical (MHD)
simulations currently favour disks with strongly changing viscosity in
radial and vertical directions
\citep[e.g.][]{Gammie1996,Turner2007,Bai2013}. In these simulations
turbulence is only expected to be high at those locations where the
ionisation fraction of the gas is also high, that is, the upper and outer
regions of the disk. In the disk mid-plane the viscous $\alpha$ may be
as low as 10$^{-6}$, especially if non-ideal MHD effects are taken
into account. In this regime, radial and vertical mixing due to the
turbulence becomes a very slow process, with mixing time-scales close
to the disk lifetime. Laminar accretion through the midplane would still enrich the inner disk 
midplane with \ce{CO2} but the low vertical mixing would greatly 
reduce the amount of material that is lifted into the IR-emitting 
layers of the disk. A mixing time of $10^6$ years would imply an 
effective vertical $\alpha$ of $10^{-6}$ for a \ce{CO2} iceline at 
10 AU (Eq.~\ref{eq:vertmixing}, \citealt{Ciesla2010}). Note that 
this still means that \ce{CO2} is mixed upwards within $10^5$ years
in the inner 1 AU, which would be visible as an added warmer 
component to the \ce{CO2} emission. Observations of another 
species created primarily on grains at high abundance, $ > 10^{-7}$,  
such as \ce{CH4}, \ce{NH3} or \ce{CH3OH} near their respective icelines 
can be used to rule out inefficient vertical mixing.

\subsubsection{Trapping of {\ce{CO2}} in large solids}
\label{sss:trapping}
There are ways to trap a volatile in a less volatile substance and
prevent or delay its sublimation to higher temperatures. One
possibility is to lock up the \ce{CO2} inside the water ice. This can
be done in the bulk of the amorphous ice, below the surface layers, or
by the formation of clathrates in the water ice at the high midplane
pressures. Even in the case that \ce{CO2} is completely trapped in
\ce{H2O} ice, \ce{CO2} still comes off the grain together with the
\ce{H2O} at the \ce{H2O} iceline. At this point the \ce{CO2} is in the
gas phase and free to diffuse around in the disk. The strong negative
abundance gradient outside the \ce{H2O} iceline will transport the
\ce{CO2} outwards until the \ce{CO2} starts to freeze-out of its own
accord at the \ce{CO2} iceline. This would still result in gas-phase
abundances within the \ce{CO2} iceline close to the original icy \ce{CO2}
abundances of the outer disk. 

Spherical grains between 0.1 $\mu$m and 1 mm are coated
with $10^2$ to $10^{10}$ layers of water ice, respectively. In the
case of small particles, with only 100 layers of ice, it is hard to
imagine that the water traps a significant amount of \ce{CO2}. For the
larger grains with a smaller relative surface area, however, the water ice could
readily trap a lot of \ce{CO2}. In a fully mixed \ce{H2O} and \ce{CO2} ice,
the sublimation of \ce{CO2} from the upper 100 million ice layers will 
still keep the fraction of sublimated \ce{CO2} below 1\%, however efficient grain fragmentation, such as assumed in our models, will expose \ce{CO2} rich layers allowing for the sublimation of more \ce{CO2}. For
small grains, clathrates may keep the \ce{CO2} locked-up in the upper
layers to prevent sublimation. However, to trap a single \ce{CO2}
molecule at least six \ce{H2O} molecules are needed
\citep{Fleyfel1991}. 

For amorphous ices similar restrictions exists on the \ce{H2O} to
\ce{CO2} ratio for trapping \ce{CO2} in the water ice. For ices with a
\ce{H2O} to \ce{CO2} ratio of 5:1, \ce{CO2} desorption happens at the
\ce{CO2} desorption temperature, only at a ratio of 20:1 is the
\ce{CO2} fully trapped within the \ce{H2O} ice
\citep{Sandford1990}. \cite{Collings2004} show that the majority of the
trapped \ce{CO2} desorps at temperatures just below the desorption
temperature of \ce{H2O}, due to the crystallization of
water. \cite{Fayolle2011} note that the fraction of trapped \ce{CO2}
also depends on the thickness of the ice. They are able to trap
65\% of the \ce{CO2} in a 5:1 ice, which means that a significant
fraction of the \ce{CO2} still comes of at the \ce{CO2} iceline. 

Both of these mechanisms need to have all the \ce{CO2} mixed with
water. For interstellar dust grains there is observational evidence
that this is not the case: high quality {\it Spitzer} spectra indicate
that \ce{CO2} is mixed in both the \ce{CO}-rich layers of the ice as
well as the \ce{H2O}-rich layers of the ice
\citep{Pontoppidan2008}. Sublimation of \ce{CO} from these mixed ices
would create a \ce{CO2} rich layer around the mixed \ce{H2O}:\ce{CO2}
inner layers.

Another option would be to lock the \ce{CO2} in the refractory
material. We note however that the \ce{CO2} ice abundance is of the same
order as the \ce{Mg}, \ce{Fe} and \ce{Si} abundances
\citep{Asplund2009}, as such it is unlikely that \ce{CO2} can be
locked up in the refractory material without chemical altering that
material, such as conversion into, for example, \ce{CaCO3}.

Measurements from 67P/Churyumov-Gerasimenko indicate that even a
kilometre-sized object outgasses \ce{CO2} at 3.5 AU distance
\citep{Hassig2015}. For larger objects, \ce{CO2} retention might be
possible, but locking up 99\% of the \ce{CO2} ice in these bodies
seems highly unlikely.

\subsubsection{Trapping icy grains outside of the iceline}
\label{sss:OutsideIceline}
It is possible to stop the \ce{CO2} from sublimating if the \ce{CO2}
ice never reaches the \ce{CO2} iceline, for example by introducing a
dust trap \citep{vanderMarel2013}. Stopping the \ce{CO2} ice from
crossing the iceline would also mean stopping the small dust from
crossing the iceline. This would lead to fast depletion of the dust in
the inner disk, creating a transition disk, although some small dust
grains can still cross the gap \citep{Pinilla2016}. Indeed, most disks
for which \ce{CO2} has been measured have strong near- and
mid-infrared continuum emission, indicating that the inner dust disks
cannot be strongly depleted. Furthermore, measurements by
\cite{Kama2015} show that for Herbig Ae stars, the accreting material
coming from a full disk has a refractory material content close to the
assumed ISM values even though these disks have no detectable infrared
\ce{CO2} emission with {\it Spitzer} \citep{Pontoppidan2010} .

\subsubsection{Shocks}

For high-mass protostars, \cite{Charnley2000} proposed that C-shocks
with speeds above 30 km s$^{-1}$ passing through the region within the
\ce{CO2} iceline could account for the low observed \ce{CO2}
abundance. Dense C-shocks can elevate temperatures above 1000 K and
create free atoms such as \ce{H}, \ce{N} and \ce{C}, so all of the
\ce{CO2} in the shocked layer can then be efficiently converted to
\ce{H2O} and \ce{CO}. At the higher densities in disks, lower velocity
shocks can reach the same temperature.  For example,
\cite{Stammler2014} calculate that a 5 km s$^{-1}$ shock would reach
temperatures upwards of 1000 K in the post shock gas.  If shocks with
such speeds occur in disks as well this could be an explanation for
the low \ce{CO2} abundance. However, assuming the shock only destroys
gas-phase \ce{CO2}, not the ice, shocks must happen at least once
every $3\times 10^3$ years ($1/10^{-11}$~s$^{-1}$) and the shock
speeds have to be of the order of the Keplerian velocity.

These nebular shocks have been invoked to explain the ubiquitous
chondrules found in meteorites. For these chondrules to form, dense
gas needs to be flash heated, so that grains can evaporate and quickly
reform as the gas cools \citep{Hewins2005}. \cite{Desch2002} proposed
that a slow shock (~7 km s$^{-1}$) in the dense midplane can heat up
the gas to temperature around 2000 K, enough to start sublimation (see
also \citealt{Stammler2014}). In the case of efficient cooling, it
takes the gas a few minutes to cool, but this would be enough for the
\ce{CO2 + H -> CO + OH} reaction to significantly reduce the \ce{CO2}
abundance. Slower shocks, with low cooling rates would also be able to
destroy \ce{CO2} efficiently, as long as temperatures of $750$ K are
reached (see Fig.~\ref{fig:CO2_temp}). To effectively destroy \ce{CO2},
these shocks need to affect the entire inner disk at least once every
few thousand years. Each shock could then also induce chondrule
formation.  There is evidence that chondrules went through multiple
creation events, but the frequency of these events is currently not
constrained \citep{Ruzicka2008,Desch2012}.

Further out in the disk, shocks can sputter ices, including \ce{CO2}
ice, from the grains
\citep{Charnley2000}. However, sputtering does not necessarily destroy
the \ce{CO2}.  Moreover, full destruction of ices by shocks is
unlikely since \ce{CO2} needs to survive in the comet formation
regions of the disk.

\subsubsection{UV-dominated layers}
\label{sss:UV}
\begin{figure}
\includegraphics[width=\hsize]{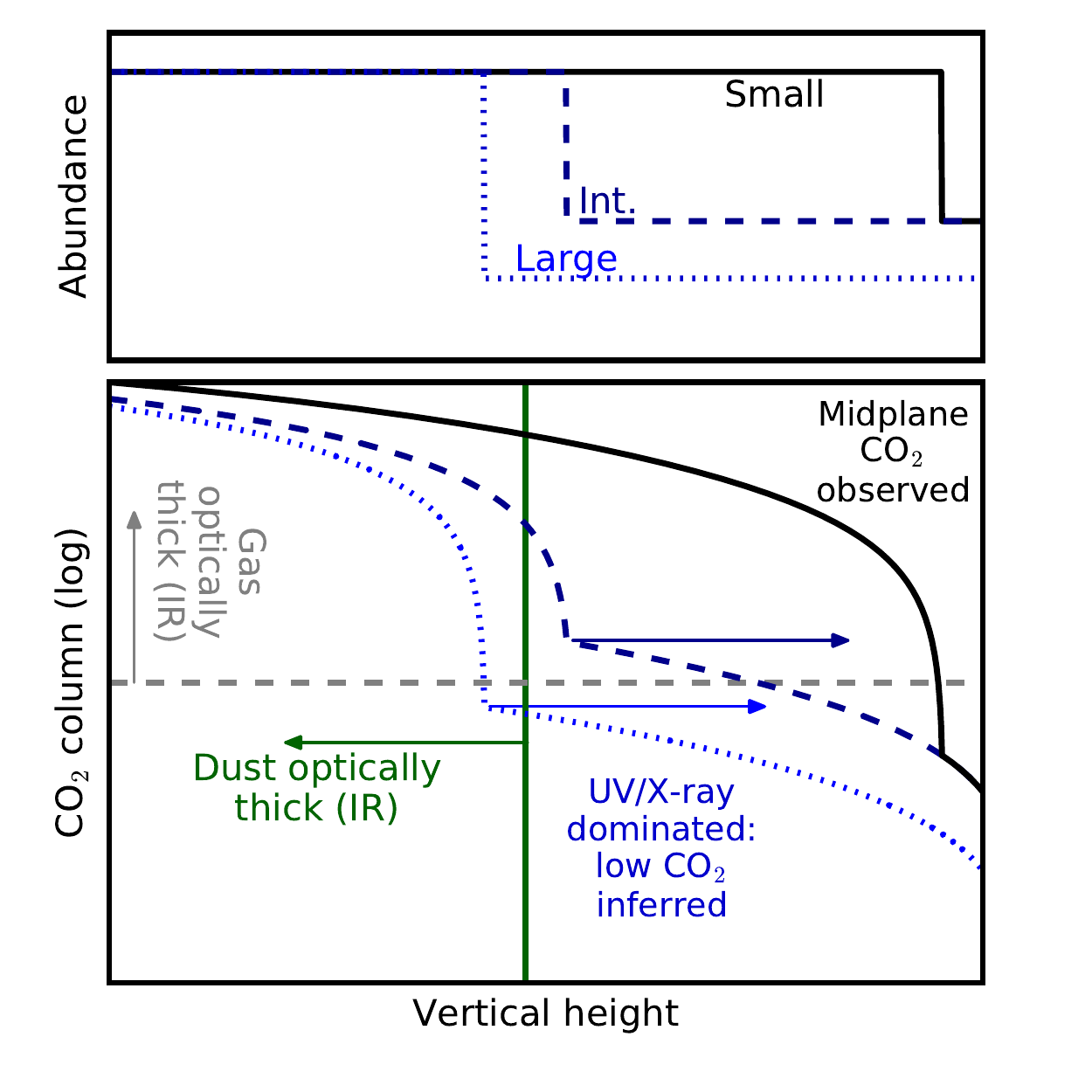}
\caption{\label{fig:penetration} Schematic of the vertical \ce{CO2} abundance (top panel) and upwards column (bottom panel) as function of disk height. Three different cases are shown in solid black, dashed dark blue and dotted blue lines. The black line shows the case for shallow UV or X-ray penetration, in this case an abundance similar to the midplane \ce{CO2} abundance will be measured. The dashed dark blue line shows a slightly deeper penetration of radiation, in which case a lower \ce{CO2} abundance will be inferred from the observations because the gas is optically thick and hiding the higher abundance deeper into the disk. The dotted blue line shows the case of deep penetration. In this case a low \ce{CO2} abundance will be inferred due to the dust hiding the bulk of the \ce{CO2}. The vertical lines in the upper panel show where the disk becomes optically thick to \ce{CO2} lines, the solid and dashed grey line, show where the gas becomes optically thick for the small and intermediate PDR or XDR respectively, the green dotted line shows where the dust becomes optically thick, which is above where the gas becomes optically thick for the large PDR or XDR. }
\end{figure}

\begin{figure*}
\includegraphics[width = \hsize]{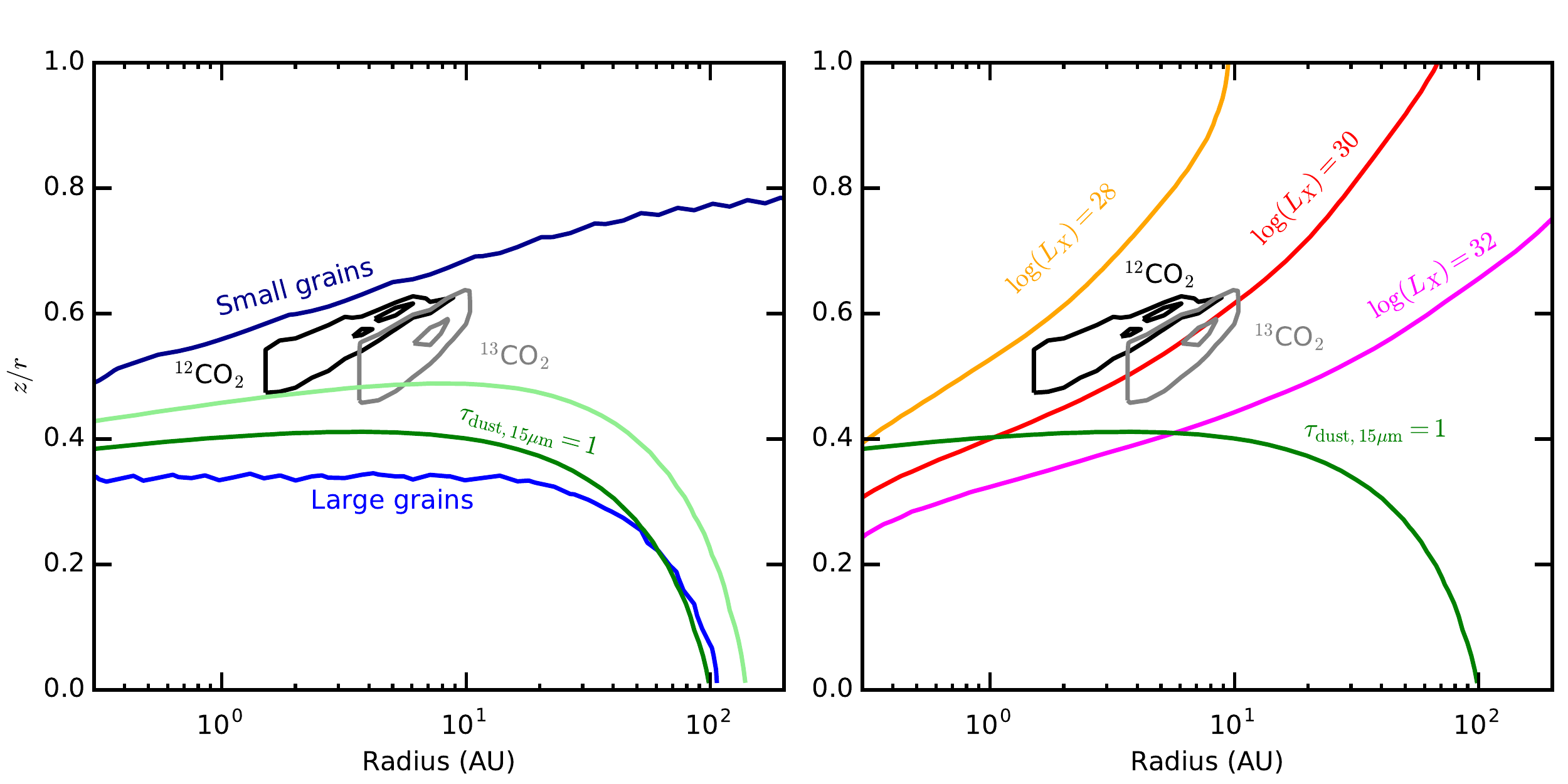}
\caption{\label{fig:UV/XRAYdisk} 
\emph{Left:} depth into the disk to which UV can destroy \ce{CO2} at a rate of $10^{-11}$ s$^{-1}$ for a disk with an upper disk gas-to-dust ratio of 1000 and large grains (cyan) and small grains (blue). The black and grey contours show the area from which 25\% and 75\% of the emission of \ce{^{12}CO2} and \ce{^{13}CO2} respectively originates. The green lines show the dust $\tau = 1$ surface at 15 $\mu$m for the large grains (up to 1 mm, dark green) and small grains (up to 1 $\mu$m, light green) respectively. The UV photo destruction rate has a stronger dependence on the grain size distribution than this $\mu$m dust photosphere. 
\emph{Right:} depth into the disk to which X-rays can destroy \ce{CO2} at a rate of $10^{-11}$ s$^{-1}$ for different stellar X-ray luminosities ($L_X$ in erg s$^{-1}$). The black and grey contours show the area from which 25\% and 75\% of the emission of \ce{^{12}CO2} and \ce{^{13}CO2} respectively is coming from. The green line shows the dust $\tau = 1$ surface at 15 $\mu$m for the large grains (up to 1 mm).}
\end{figure*}

The inner disk chemical models by \cite{Walsh2015} show that there is
a surface layer in the disk where UV destruction becomes important
resulting in abundances of $10^{-8}$--$10^{-7}$. Depending on the
time-scale of the chemical processes and the extent of this layer, it
may be optically thick in the \ce{^{12}CO2} infrared lines. This means
that low abudances of \ce{CO2} will be measured in \ce{^{12}CO2}
irrespective of the amount of sublimating \ce{CO2} ice in the
midplane. If this UV-dominated layer is not too thick, \ce{^{13}CO2}
could still have an imprint of the deeper, more abundant \ce{CO2} that
is coming from the iceline. On the other hand, if this layer is thick
enough that even \ce{^{13}CO2} becomes optically thick to its own
radiation, then the more abundant \ce{CO2} sublimating from the
iceline would be hidden from our view (see 
Fig.~\ref{fig:DALI_fluxratios} and Fig.~\ref{fig:penetration}). This would happen at \ce{CO2} columns
larger than $10^{18}$ cm$^{-2}$.  

The (vertical) thickness of the photon dominated layer is very
dependent on the UV flux and dust opacity in the upper layers of the
disk. For a disk that has only small grains ($< 1\mu$m) in the upper atmosphere, 
the UV opacity is high. In this case the UV dominated layer only contains a short 
column of gas, and thus \ce{CO2} (Fig.~\ref{fig:UV/XRAYdisk}). Observations of 
such a disk would lead to the spectra as presented in Sect.~\ref{ssc:Dalimodels}.
If it is assumed that grains have grown to mm sizes, even in the upper atmosphere, 
then the UV can penetrate a lot deeper. This leads to a larger column of UV dominated gas
and that could hide the high abundance \ce{CO2} in the midplane, if the \ce{CO2} 
in this layer becomes optically thick to its own IR radiation or if the dust becomes
optically thick at IR wavelengths in the UV dominated layer. The first scenario is sketched 
in panel C of Fig.~\ref{fig:schematic}. To
probe the midplane \ce{CO2} abundance one would need to observe disks with a lot 
of small grains in the upper atmosphere, preferably around UV weak sources such that
the \ce{CO2} column in the UV dominated layer is minimised.

\subsubsection{X-ray induced dissociation}
\label{sss:X-ray}
T-Tauri stars are known to have strong X-ray emission
$10^{28}$--$10^{34}$ ergs\,s$^{-1}$ \citep{Feigelson2002}. These
X-rays can ionise the gas and create a local UV field in the same way
as cosmic-rays, destroying molecules in the gas \citep[see, for
example, the models by ][ for high mass protostars]{Stauber2005}. As
X-rays are more quickly attenuated than the higher energy cosmic-rays,
they are not efficient in destroying \ce{CO2} in the disk midplane,
however, X-rays do penetrate further into the disk atmosphere than UV
photons. Thus, X-rays can help in decreasing the \ce{CO2} abundance in the
atmosphere of the disk, especially if hard X-ray flare, such are
thought to occur in TW Hya are common in other T-Tauri stars as well
\citep{Kastner2002,Cleeves2015,Cleeves2017}.  

For our specific DALI model, a total X-ray flux of
$10^{30}$~erg~s$^{-1}$, emitted as a $10^7$ K black body, would have a
destruction rate of $10^{-11}$ s$^{-1}$ in most of the \ce{^{12}CO2}
emitting region and could thus have an effect on the \ce{^{12}CO2}
fluxes (Fig.~\ref{fig:UV/XRAYdisk}). A total X-ray flux of
$10^{32}$\,erg\,s$^{-1}$ would be needed to have the same destruction
rate in most of the \ce{^{13}CO2} emitting region. Thus, a source that
has a time-averaged X-ray luminosity $>10^{30}$~erg~s$^{-1}$ could
show a low \ce{CO2} abundance in the atmosphere and have a higher
abundance underneath. If the X-ray luminosity of the source is lower
than $10^{32}$\,erg\,s$^{-1}$, this should be reflected in the
\ce{^{13}CO2} measurement. Note however that the X-rays do not reach
into the midplane mass reservoir. Since the vertical transport for a
fully viscous disk is orders of magnitude faster than the radial
transport, the replenishment rate of \ce{CO2} in the upper layers due
to mixing from the midplane would also be faster. Thus, these X-ray
luminosities should be considered as lower limits. Most disk hosting stars 
do not have high enough X-ray luminosities to destroy \ce{CO2} 
efficiently \citep{Bustamante2016}, but X-ray destruction of \ce{CO2} in the 
surface layers can be an explanation in some sources and lead to similar 
effects as for the case of UV, such as the \ce{CO2} IR lines becoming
optically thick in the X-ray dominated region.

\section{Summary and conclusions}
We have presented a model for the transport of major gas and ice species.
This model predicts that transport of ices from the outer disk towards 
the inner disk can have a significant effect on the abundance of species 
in the inner disk. Radial transport is predicted to be so efficient, that 
even in the case without any radial drift, the modelled \ce{CO2} abundance 
in the inner disk is orders of magnitude higher than the currently available 
observations of disks. As such \ce{CO2} cannot be directly used to trace the mass transport rate. The presence of dust in disks older than a Myr does imply that the mass transport rate of dust cannot be larger than $ 10 \times \dot{M}_\mathrm{acc,gas} / \left(\mathrm{g/d}\right)$. This is inconsistent with the pebble-sized dust observed in disks.

To reconcile model and observation \ce{CO2} abundances, either the midplane \ce{CO2} needs to be destroyed by some physical or
chemical mechanism, or the \ce{CO2} abundance in the upper layers needs to be decoupled from the \ce{CO2} abundance in the midplane. This can be achieved by preventing vertical \ce{CO2} cycling, hiding the \ce{CO2} under the dust IR photosphere or by the creating a low abundance \ce{CO2} column that is optically thick in a UV or X-ray dominated layer. 
We can summarize this work with the following
conclusions.

\begin{itemize}

\item Accretion flow and diffusion equilibrate the inner disk \ce{CO2}
  gas abundance with the high outer disk \ce{CO2} ice abundance of
  $\sim 10^{-5}$ in less than 1 Myr for $\alpha >10^{-3}$
  (Sect.~\ref{ssc:PureViscous}).

\item For any disk that retains its dust on the same time-scale as it
  retains the gas, dust diffusion processes dominate over radial drift
  (Sect.~\ref{ssc:ViscousandGrowth}~and~App.~\ref{app:growthmodels}).

\item If dust can grow efficiently and drift inwards, both \ce{CO2} and
  \ce{H2O} will be enhanced above the outer disk ice abundance as long
  as the dust is drifting inwards with a mass flux more than 1\% of
  that of the gas,
  $\dot{\mathrm{M}}_\mathrm{dust} \gg 0.01\,\dot{\mathrm{M}}_\mathrm{gas}$. 
  (Sect.~\ref{ssc:Total_viscous})
  
\item A physical or chemical process that would induce a vertically
  averaged \ce{CO2} destruction rate of at least $10^{-11}$ s$^{-1}$
  ($<10^3$ yr) is necessary to produce \ce{^12CO2} fluxes that are
  consistent with the current observations of disks.
(Sect.~\ref{ssc:Dalimodels}).

\item Within the current chemical networks, there are no paths to
  destroy \ce{CO2} fast enough without invoking strong X-ray or
  UV-fluxes or elevated cosmic ray fluxes. Especially at temperatures between
  60 and 150 K, destroying \ce{CO2} is extremely
  difficult in the current chemical networks since the formation of
  \ce{CO2} is strongly favoured over the production of \ce{H2O} in
  that temperature range. Above 300 K \ce{CO2} destruction pathways exists, 
  but they only become efficient enough at temperatures above 750 K. (Sect.~\ref{ssc:FormDestr},~\ref{ssc:Chemdisc}~and~App.~\ref{app:chem}).

\item Shocks could raise temperatures enough to destroy \ce{CO2} efficiently,
  but effectively lowering the \ce{CO2}
  abundance would require the entire inner 10 AU of the disk to be
  processed by a shock every $10^3$ years.

\item Trapping \ce{CO2} in water ice seems unlikely and is not
  expected to lower the inner disk gaseous \ce{CO2} abundance
  significantly. Trapping the grains before they cross the iceline
  efficiently enough to lower \ce{CO2} abundances in the inner disk
  would quickly create a transition disk
  (Sect.~\ref{sss:trapping}~and~\ref{sss:OutsideIceline}).

\item Deep penetration of UV and strong X-ray fluxes can lower the
  \ce{CO2} abundance in the \ce{^{12}CO2} emitting region. This can
  happen if grains in the disk atmosphere have grown to mm sizes and
  the central source has enough UV luminosity ($> 10^{27}$ erg s$^{-1}$, about 0.1\% of the solar UV).
  Likewise sources that have an
  X-ray luminosity greater than $10^{30}$ erg s$^{-1}$ efficiently
  destroy \ce{CO2} in a large part of the IR emitting region. If the UV or X-ray dominated region does not reach down to the dust 15 $\mu$m $\tau = 1$ surface, \ce{^{13}CO2} can still contain a information on the amount of sublimated \ce{CO2}.

\item A high ratio of the \ce{^{13}CO2} $Q-$branch flux over the
  \ce{^{12}CO2} $Q-$branch flux $(>0.05)$, can hint at an contribution
  of sublimating \ce{CO2} at the iceline. Sublimating \ce{CO2} is,
  however, not the only way to get high flux ratios. A more
  complete, source specific analysis would be needed to exclude other
  possibilities (Sect.~\ref{ssc:Dalimodels}).

\end{itemize}

In summary, to explain the discrepancy between our models and observations, we either need to invoke frequent shocks in the inner 10 AU, or need to hide the abundant midplane \ce{CO2} from our view. This can be due to very inefficient mixing or due to deep UV and/or X-ray penetration. In the case of UV penetration, this would imply efficient grain growth and settling, in the case of X-rays, this would be stronger than measured X-ray fluxes for most sources, or frequent X-ray flares. The key to hiding the mid-plane \ce{CO2} is that the low abundance \ce{CO2} in the upper layers has a high enough column density to become optically thick to its own IR radiation. Trapping the \ce{CO2} on the grains, or destroying it in the gas phase near the iceline at ambient temperatures is strongly disfavoured. 

Is \ce{CO2} a special case in terms of discrepancy between models and
observations?  Modelling and observations of other molecules abundant
in ices, such as \ce{NH3} and \ce{CH4}, may be used to test whether
a chemical or physical reason is the origin of the current
discrepancy.  Ultimately, such data can constrain the mass flow of
solids if the chemical composition of the surface layers is
sufficiently representative of that in the midplane.

\section*{Acknowledgements}
Astrochemistry in Leiden is supported by the European Union A-ERC
grant 291141 CHEMPLAN, by the Netherlands Research School for
Astronomy (NOVA), by a Royal Netherlands Academy of Arts and Sciences
(KNAW) professor prize. Daniel Harsono, Stefano Facchini, Kees
Dullemond and Til Birnstiel are thanked for insightful discussions and 
Catherine Walsh and Christian Eistrup for sharing and help with the
gas-grain chemical network.
   
\bibliographystyle{aa}
\bibliography{../../Literature/Lit_list}

\begin{appendix}
\section{UV dust cross sections}
\begin{figure}
\centering
\includegraphics[width=\hsize]{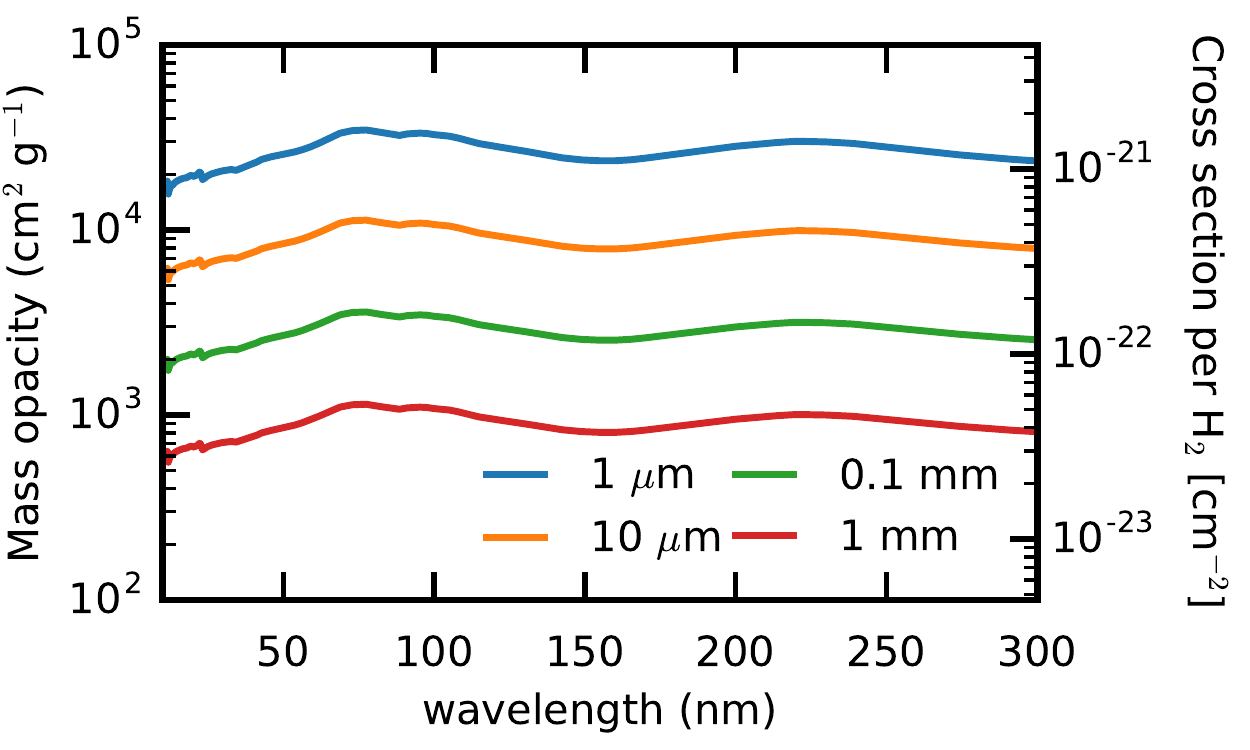}
\caption{\label{fig:opacities} UV Opacities and cross section of the dust for different sizes of the largest grains. The MRN power law slope is 3.5 for all cases. Growing grains from 1 $\mu$m to 1 mm has more than an order of magnitude effect on the opacities.}
\end{figure}

Fig.~\ref{fig:opacities} shows the UV opacities and cross sections for different grain size distributions used in Sect.~\ref{ssc:CRdestr}.

\section{Chemical modelling}
\label{app:chem}
\subsection{Gas-phase only models}
\label{app:gasmodel}
To test the main \ce{CO2} destruction and formation processes, a small grid of chemical models was run. The network from \cite{Walsh2014} was used as basis and modified. The gas-phase chemical network is based on the UMIST 2012 network \citep{McElroy2013}. The network includes freeze-out and desorption. \ce{H2} formation on grains following \cite{Cazaux2004}, which includes \ce{H2} formation at high temperatures. The grain-surface reactions included in the network were removed for this test. As the main interest is in the midplane layers, the local UV and X-ray flux are assumed to be negligible. The cosmic ray rate is taken to be 10$^{-17}$ s$^{-1}$. 

Four series of models were run, two starting from `atomic' initial conditions (all species are in atomic form except for \ce{H2}, the other two from `molecular' conditions (see Table~\ref{tab:initialchem}, the atomic initial conditions use the same elemental abundances as the used for the molecular test). For each set of initial conditions one set of models was run with \ce{H2} formation happening only on grains according to \cite{Cazaux2004} leading to an atomic \ce{H} abundance of $10^{-4}$ with respect to H$_2$, while in the other set of models the additional \ce{H2} formation route at high temperature was added to keep the \ce{H} abundance at $10^{-12}$ (1 cm$^{-3}$). The density was kept constant at $10^{12}$ cm$^{-3}$ and the temperature was varied between 150 and 800 K. 

\begin{figure}
\centering
\includegraphics[width=\hsize]{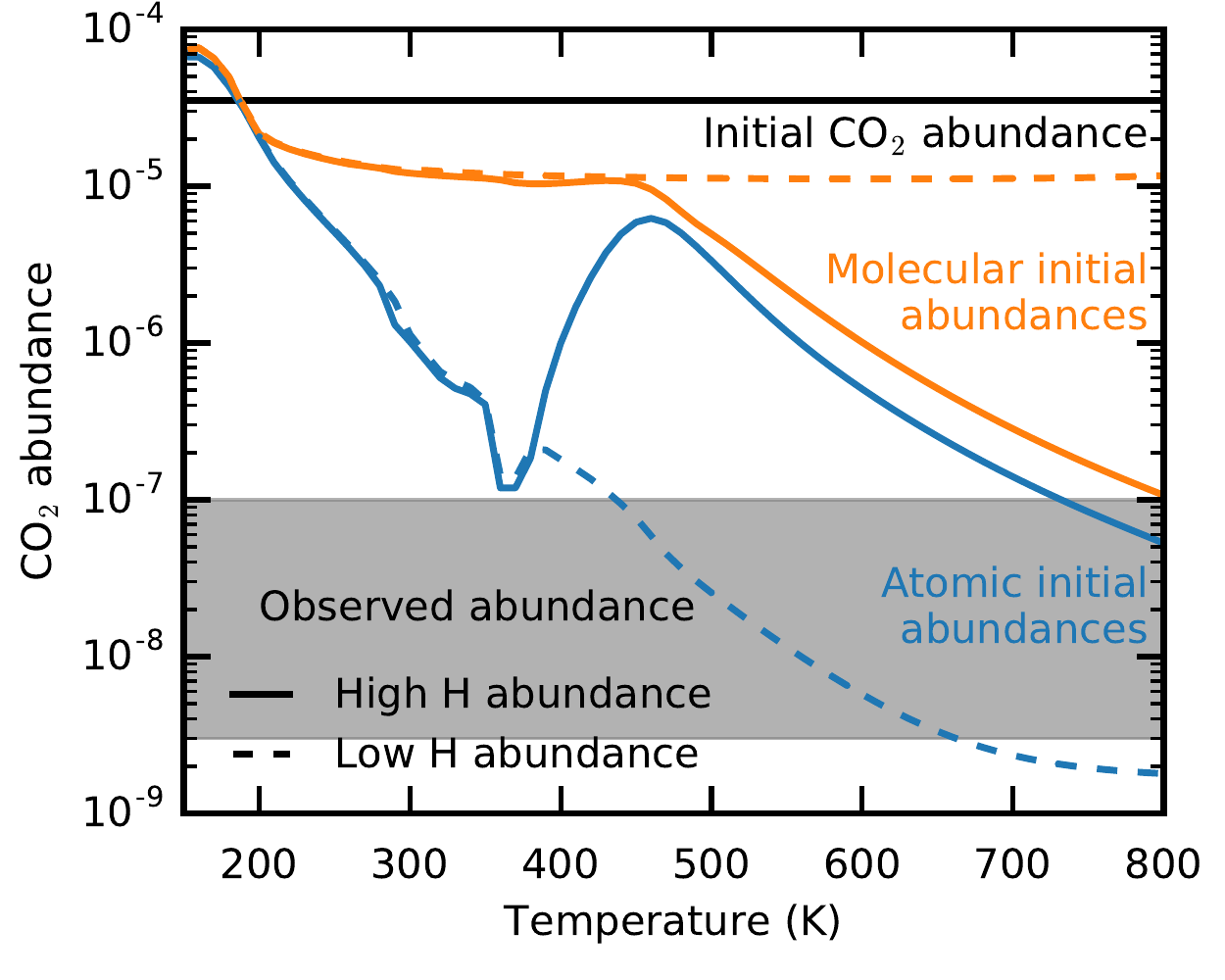}
\caption{\label{fig:CO2_temp} \ce{CO2} abundance from a gas-phase chemical network as function of temperature after 1 Myr of evolution. The density is $10^{12}$ cm$^{-3}$. }
\end{figure}

The \ce{CO2} abundances after 1 Myr of chemical evolution are presented in Fig.~\ref{fig:CO2_temp}. From this plot it can be gathered that there is no efficient destruction mechanism in the current gas-phase network until a temperature of $\sim$450 K is reached. 

For temperatures between 150 and 450 K, the different atomic \ce{H} abundances do not significantly influence the results. The difference between the initial \ce{CO2} abundance and the \ce{CO2} abundance after 1 Myr in the molecular case can be explained by the destruction of \ce{CO2} and \ce{H2O} due to processes driven by cosmic rays, e.g. cosmic ray induced photo dissociation, and the reformation of \ce{CO2} from the fragments. 

For models starting from atomic initial conditions the \ce{CO2} content decreases with increasing temperature as expected from the formation speeds of \ce{CO2} and \ce{H2O} in the model with a low \ce{H} abundance, and this trend continues up to higher temperatures. The local minimum in the \ce{CO2} abundance at 350 K is due to reactions with atomic \ce{N} before the nitrogen is locked up in other species. When the atomic nitrogen abundance is high, \ce{CO2} reacts with \ce{N} to form \ce{CO} and \ce{NO}, \ce{NO} is then further processed to \ce{H2O}, releasing a \ce{N} in the process. This significantly slows down the build-up of \ce{CO2} when both \ce{O} and \ce{OH} are available in the gas-phase in high abundances. 

In the models with a high \ce{H} abundance, reactions with atomic hydrogen become important starting at 350 K. This process, which is slightly more efficient for \ce{H2O} than for \ce{CO2}, together with the reformation of \ce{H2O} and \ce{CO2} from the resulting \ce{OH}, sets the equilibrium at high temperature. The abundance of \ce{H2O} as function of temperature is fairly constant, at $10^{-4}$, and thus the \ce{CO2} abundance decreases towards the highest temperature, as expected from the formation rate ratio given in Fig.~\ref{fig:ratefrac}. The difference between the \ce{CO2} abundance at high temperature in the case of atomic and molecular abundance can be explained by the lower \ce{CO} abundance in the case of atomic initial conditions. 

For the models with a low \ce{H} abundance, \ce{CO2} is not efficiently destroyed by \ce{H} at high temperatures so the \ce{CO2} abundance is nearly constant as function of temperature. From this parameter exploration we can conclude that \ce{CO2} can only be effectively destroyed by radicals like \ce{H} and \ce{N} and that this is only efficient if these are available at high abundances $>10^{-6}$. For the model beginning with atomic abundances the low temperature dependence of the abundance is similar to the high \ce{H} abundance case. Above temperatures of 400 K differences start to be noticeable. The \ce{CO2} abundance does not increase with temperature as it does in the case of high \ce{H} abundance. This is due to a similar process as in the high \ce{H} model around 350 K. All the \ce{CO2} that is formed when oxygen is not locked up in water is quickly destroyed by atomic H and N, when atomic H and N have settled to low enough abundances that \ce{CO2} can survive, all the \ce{O} is already locked up in other species, mostly \ce{H2O} and \ce{CO} leaving no extra \ce{O} to form \ce{CO2}. The small amount of \ce{CO2} that is in the gas-phase is mostly created by cosmic ray induced destruction of \ce{H2O}. The decreasing abundance towards high temperatures is due to the increasingly efficient reformation of \ce{H2O} after destruction due to cosmic rays. 

\begin{table}
\caption{\label{tab:initialchem} Initial molecular abundances for the chemical models. }
\begin{tabular}{lll|lll}
\hline
\hline
\multicolumn{6}{c}{Abundances} \\
\hline
\ce{H2} & &5.0(-1) &\ce{H} & & 0.0\\
\ce{He} & &1.4(-1) &\ce{CO} & & 1.0(-4)\\
\ce{CO2} & & 3.5(-5) &\ce{N2} & & 1.1(-5)\\
\ce{N} & & 2.1(-5)& \ce{H2O} & & 1.2(-4)\\
\ce{Si} &  &4.0(-10) & \ce{H2S} & & 1.9(-9) \\
\hline
\end{tabular}
\tablefoot{
$a(b)= a\times 10^{b}$
}
\end{table}

\subsection{Grain surface chemistry between the \ce{H2O} and \ce{CO2} icelines. }

Section \ref{app:gasmodel} investigates the gas-phase chemistry of \ce{CO2} in the region within the \ce{H2O} iceline, that is, above 150 K. In this region all species will be primarily in the gas-phase and grain surface chemistry is thus expected to be less important. Between the \ce{CO2} and \ce{H2O} icelines
The efficiency of the destruction of gaseous \ce{CO2} is estimated by investigating the chemical pathways in a full chemical model. The network from \cite{Walsh2014,Eistrup2016} was used. A point model representative for the conditions between the \ce{H2O} and \ce{CO2} iceline was run. The number density used is $10^{13}$ cm$^{-3}$, the temperature $100$ K and the cosmic ray ionisation rate $10^{-17}$ s$^{-1}$. There was no external UV in our model. After running the model for 1 Myr to get equilibrium abundances of atomic and ionised species, the chemical pathways of oxygen carrying species are examined. 

For each oxygen carrying molecule that could be created from a destruction of \ce{CO2}, we define the destruction efficiency as the fraction the oxygen atoms that is incorporated into a stable molecule, for which we take that a cosmic ray induced process is needed to destroy the molecule. Here, pathways leading to incorporation of oxygen into \ce{CO} are neglected, as these pathways are dominated by the availability of atomic \ce{C}, which must originate from an other \ce{CO} molecule. Pathways towards \ce{CO} are generally very minor. Destruction efficiencies are tabulated in Table~\ref{tab:chem_eff}. 

\begin{table}
\caption{\label{tab:chem_eff} Chemical pathways from molecules possibly created by the destruction of \ce{CO2}.}
\begin{tabular}{lll}
\hline
\hline
First generation & Destruction  & Stable resulting\\
molecule & efficiency & molecules \\
\hline
\ce{CO2+}, \ce{HCO2+} & 0\% & \ce{CO2}\\
\ce{OH} & 27\% & \ce{CO2},\ce{O2},\ce{H2O2} \\
\ce{O+}, \ce{OH+}, \ce{H2O+}, \ce{H3O+} & 100\% & \ce{H2O} \\
\ce{O}, \ce{NO}, \ce{HNO} & 82\% & \ce{O2}, \ce{CO2}, \ce{H2O2} \\
\ce{SO}, \ce{SO2} & 100\%& \ce{O2} \\
\ce{OCN}& 0\%& \ce{CO2} \\
\ce{HCNO}& 100\%& \ce{O2} \\
\ce{SiO} & 100\%& \ce{SiO} \\  
\hline
\end{tabular}
\end{table}

From the efficiencies it can be seen that direct ionisation or protonation of \ce{CO2} is not an effective way to destroy it since \ce{CO2} will be reformed from the products in most cases. Only if an oxygen atom can be removed from the central carbon, is there a chance that the oxygen will be locked up into something other than \ce{CO2}. In these cases the efficiency is 27\% (for OH) or higher. 

\section{Viscous evolution and grain growth}
\label{app:growthmodels}
\begin{figure*}
\includegraphics[type=pdf,ext=.pdf,read=.pdf,width=\hsize]{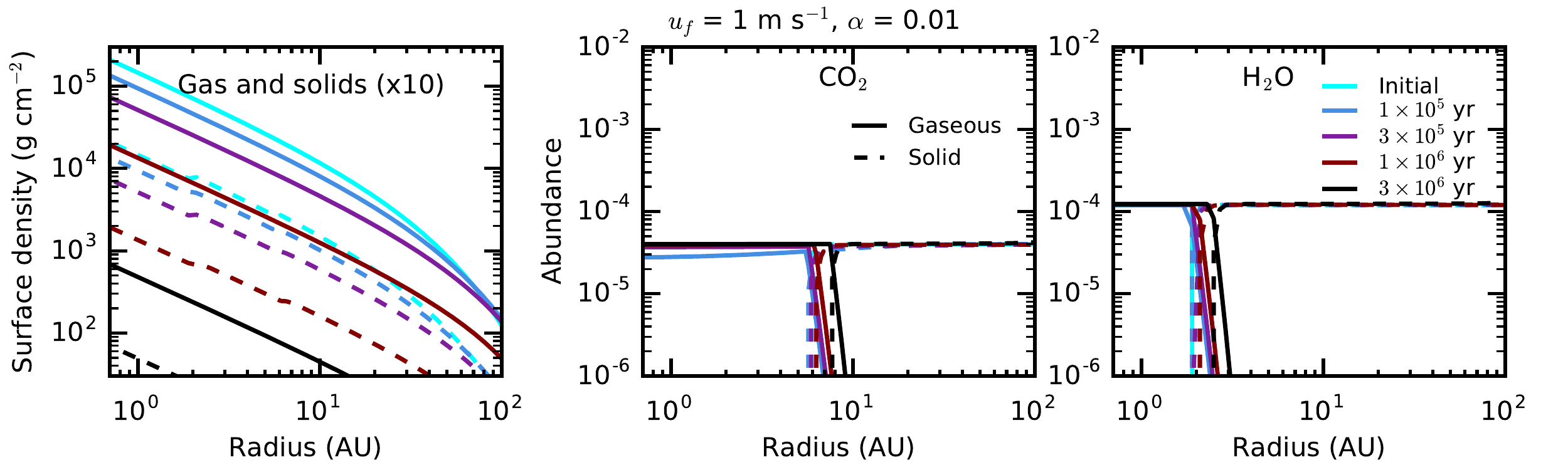}
\caption{\label{fig:vfr_1_alpha_-2}Time evolution series for a model with grain growth. This model assumes an $\alpha$ of 10$^{-2}$. The fragmentation velocity for this model is 1~m~s$^{-1}$. Panels as in Fig.~\ref{fig:nogrowthmodel}.}
\end{figure*}

\begin{figure*}
\includegraphics[type=pdf,ext=.pdf,read=.pdf,width=\hsize]{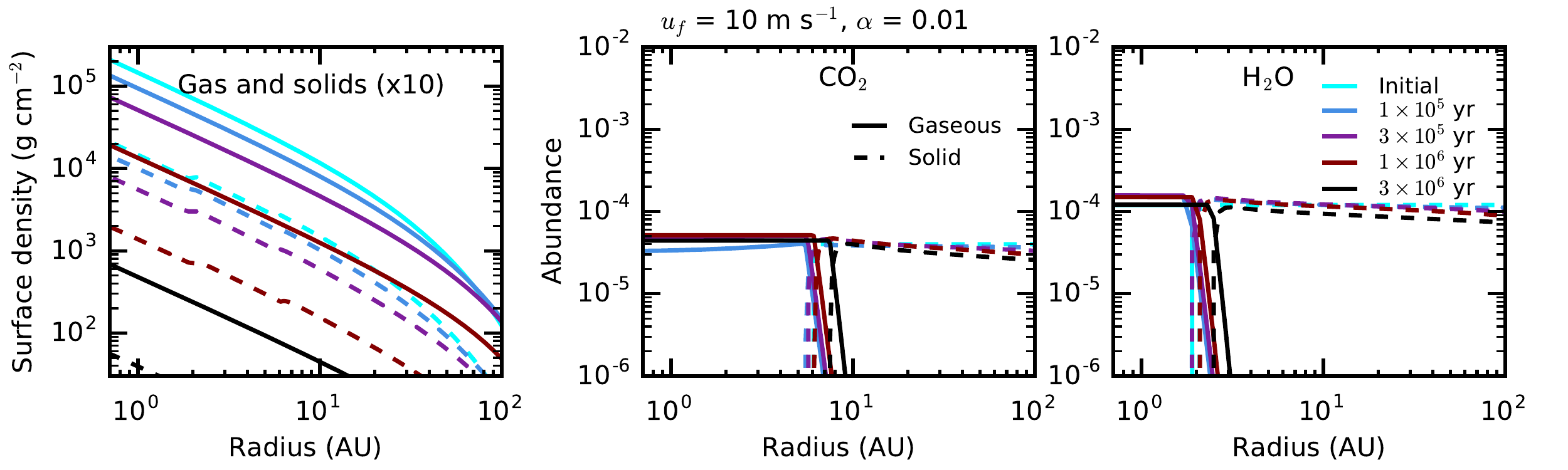}
\caption{\label{fig:vfr_10_alpha_-2}Time evolution series for a model with grain growth. This model assumes an $\alpha$ of 10$^{-2}$. The fragmentation velocity for this model is 1~m~s$^{-1}$. Panels as in Fig.~\ref{fig:nogrowthmodel}.}
\end{figure*}

\begin{figure*}
\includegraphics[type=pdf,ext=.pdf,read=.pdf,width=\hsize]{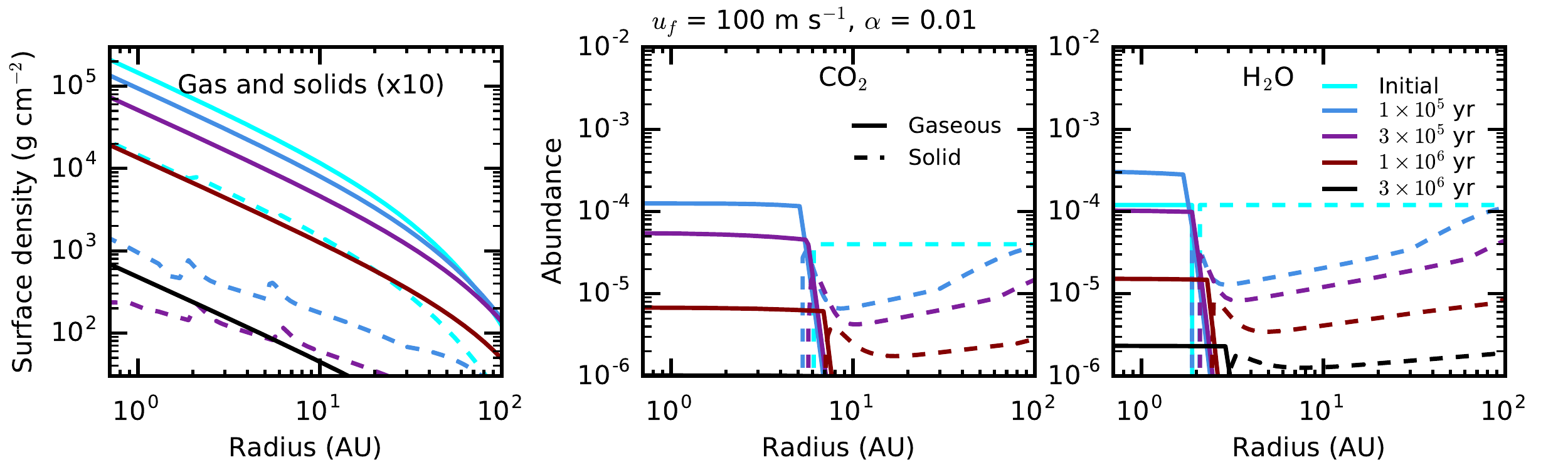}
\caption{\label{fig:vfr_100_alpha_-2}Time evolution series for a model with grain growth. This model assumes an $\alpha$ of 10$^{-2}$. The fragmentation velocity for this model is 1~m~s$^{-1}$. Panels as in Fig.~\ref{fig:nogrowthmodel}.}
\end{figure*}

\begin{figure*}
\includegraphics[type=pdf,ext=.pdf,read=.pdf,width=\hsize]{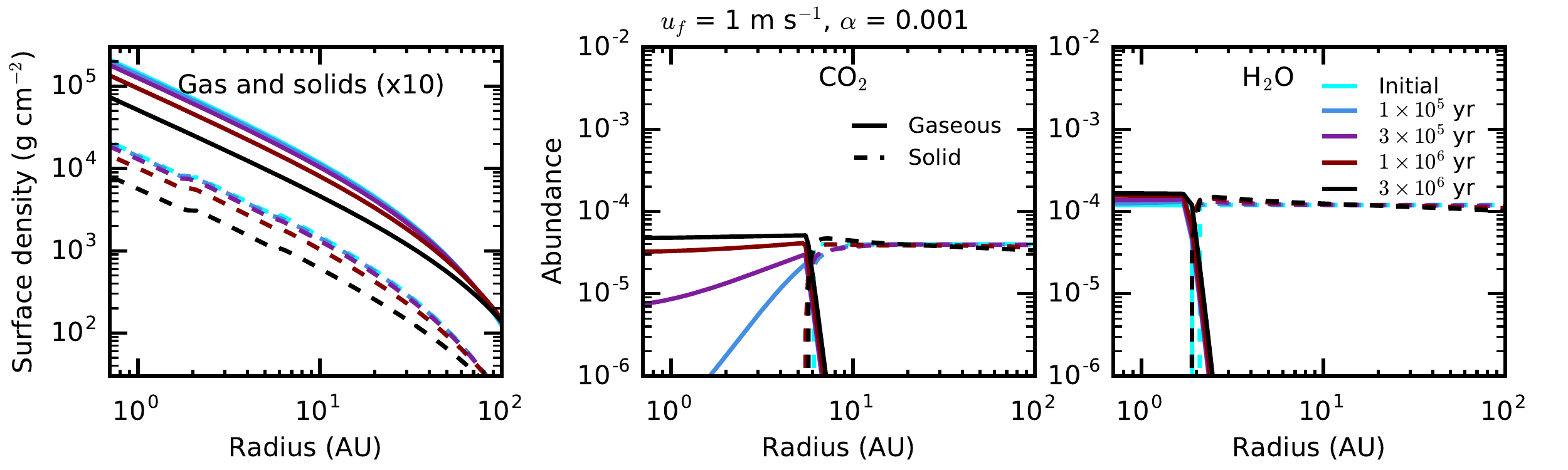}
\caption{\label{fig:vfr_1_alpha_-3}Time evolution series for a model with grain growth. This model assumes an $\alpha$ of 10$^{-3}$. The fragmentation velocity for this model is 1~m~s$^{-1}$. Panels as in Fig.~\ref{fig:nogrowthmodel}.}
\end{figure*}

\begin{figure*}
\includegraphics[type=pdf,ext=.pdf,read=.pdf,width=\hsize]{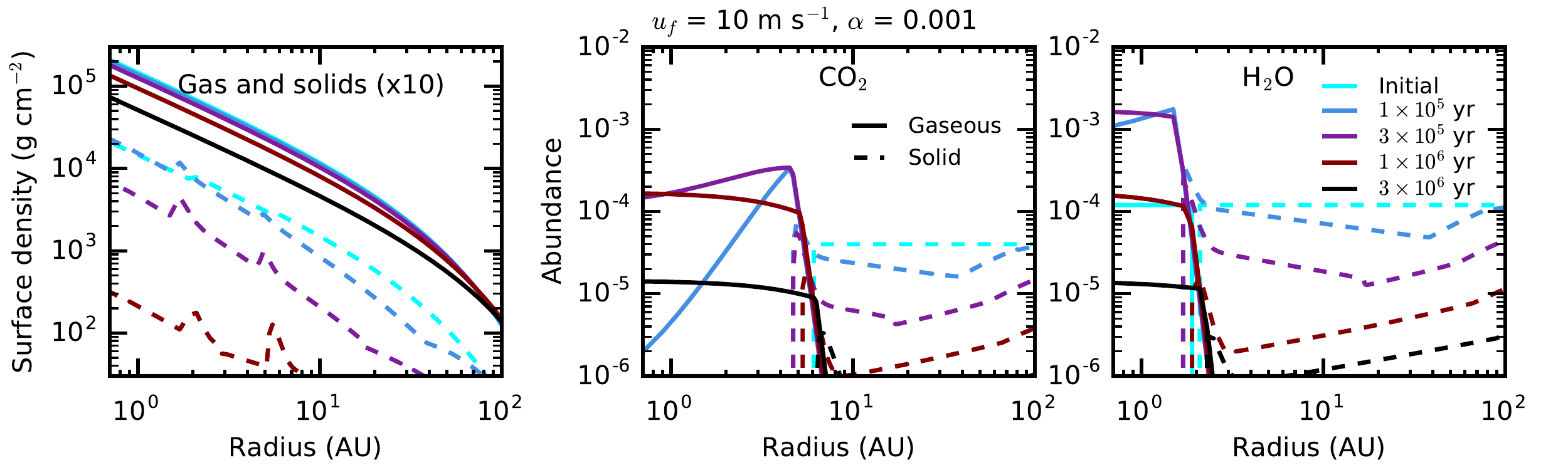}
\caption{\label{fig:vfr_10_alpha_-3}Time evolution series for a model with grain growth. This model assumes an $\alpha$ of 10$^{-3}$. The fragmentation velocity for this model is 10~m~s$^{-1}$. Panels as in Fig.~\ref{fig:nogrowthmodel}.}
\end{figure*}

\begin{figure*}
\includegraphics[type=pdf,ext=.pdf,read=.pdf,width=\hsize]{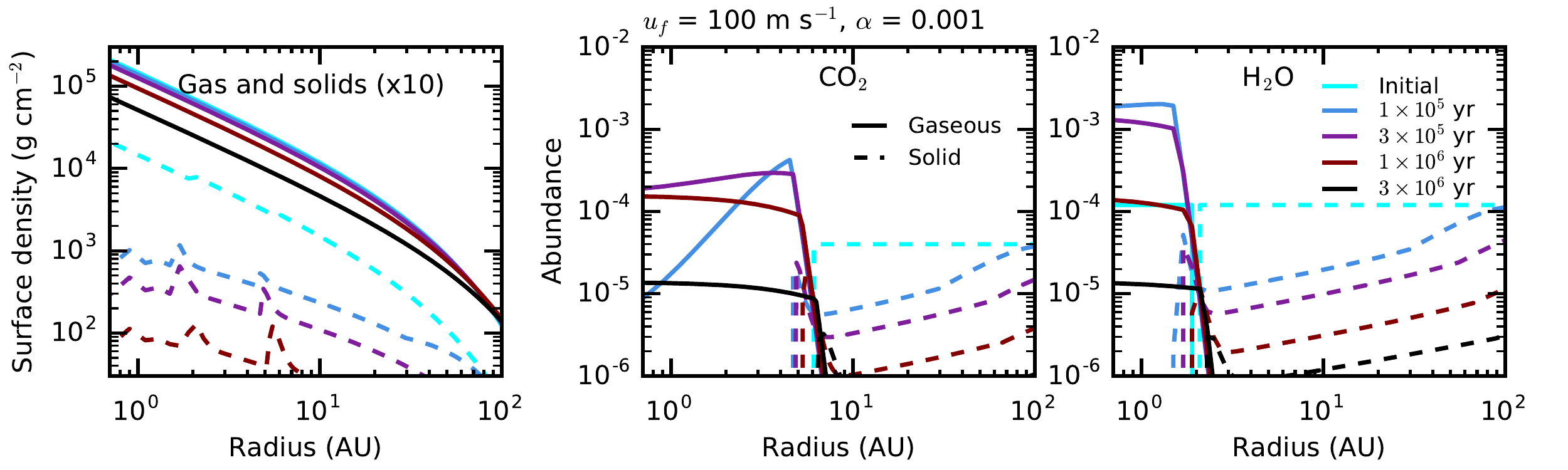}
\caption{\label{fig:vfr_100_alpha_-3}Time evolution series for a model with grain growth. This model assumes an $\alpha$ of 10$^{-3}$. The fragmentation velocity for this model is 100~m~s$^{-1}$. Panels as in Fig.~\ref{fig:nogrowthmodel}.}
\end{figure*}

\begin{figure*}
\includegraphics[type=pdf,ext=.pdf,read=.pdf,width=\hsize]{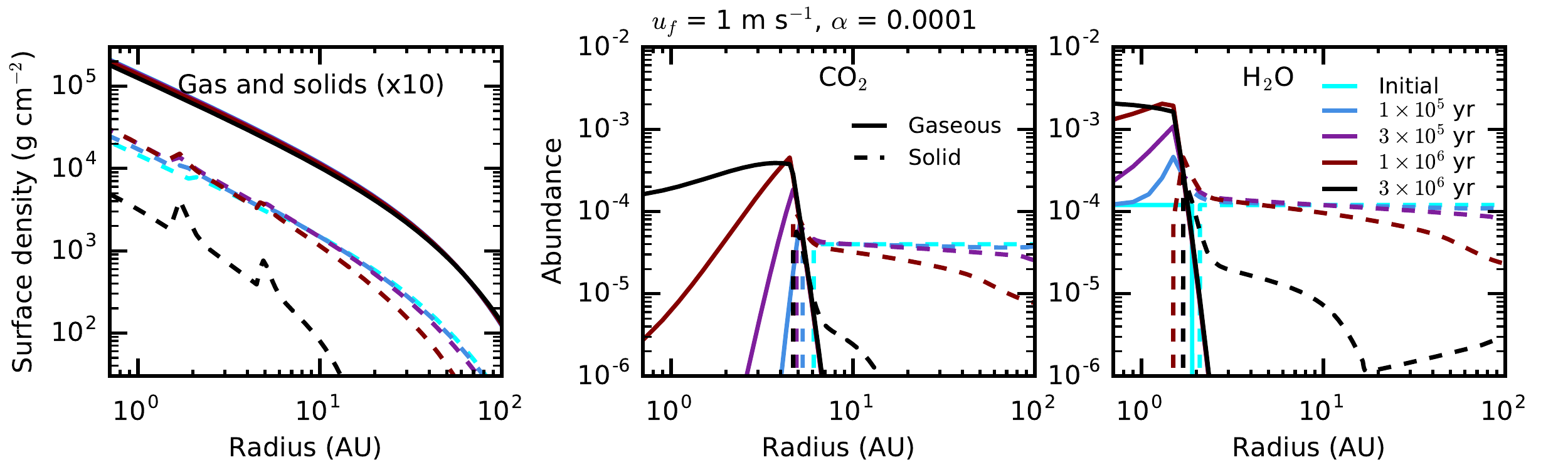}
\caption{\label{fig:vfr_1_alpha_-4}Time evolution series for a model with grain growth. This model assumes an $\alpha$ of 10$^{-4}$. The fragmentation velocity for this model is 1~m~s$^{-1}$. Panels as in Fig.~\ref{fig:nogrowthmodel}.}
\end{figure*}

\begin{figure*}
\includegraphics[type=pdf,ext=.pdf,read=.pdf,width=\hsize]{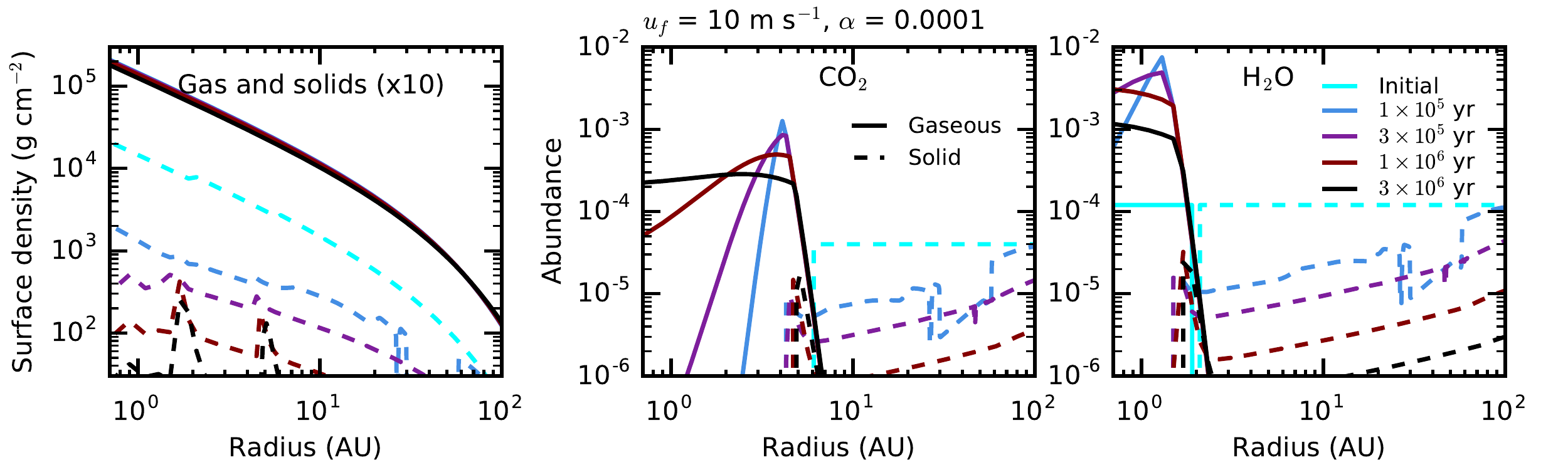}
\caption{\label{fig:vfr_10_alpha_-4}Time evolution series for a model with grain growth. This model assumes an $\alpha$ of 10$^{-4}$. The fragmentation velocity for this model is 10~m~s$^{-1}$. Panels as in Fig.~\ref{fig:nogrowthmodel}.}
\end{figure*}

\begin{figure*}
\includegraphics[type=pdf,ext=.pdf,read=.pdf,width=\hsize]{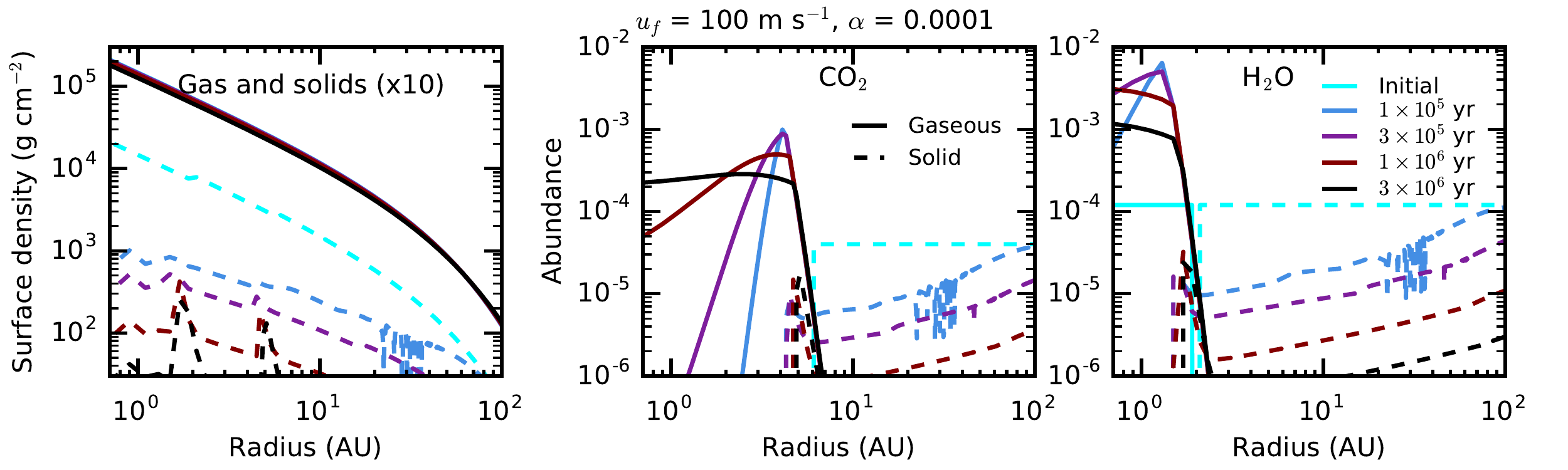}
\caption{\label{fig:vfr_100_alpha_-4}Time evolution series for a model with grain growth. This model assumes an $\alpha$ of 10$^{-4}$. The fragmentation velocity for this model is 100~m~s$^{-1}$. Panels as in Fig.~\ref{fig:nogrowthmodel}.}
\end{figure*}

Figures~\ref{fig:vfr_1_alpha_-2}~through~\ref{fig:vfr_100_alpha_-4} show model results for the model with grain growth and radial drift, but without any destruction rate. $\alpha$ values of $10^{-2}$, $10^{-3}$ and $10^{-4}$ are shown, fragmentation velocities of 1, 10 and 100 m~s${^{-1}}$ are used. The features seen in ice abundance for the higher fragmentation velocity models is due to a numerical instability in the grain growth algorithm, this does not influence our results.  

\end{appendix}

\end{document}